\documentclass[11pt]{article}

\usepackage{txfonts}

\usepackage[letterpaper,margin=1in]{geometry}

\usepackage{authblk}

\usepackage[numbers,sort&compress]{natbib}

\usepackage{graphicx}
\usepackage{array}
\usepackage{placeins}
\usepackage[font=small,labelfont=bf]{caption}

\usepackage{framed,multirow}

\usepackage{amssymb}
\usepackage{latexsym}
\usepackage{amsmath}

\usepackage{url}
\usepackage{xcolor}
\definecolor{newcolor}{rgb}{.8,.349,.1}

\usepackage[colorlinks=true,linkcolor=blue,citecolor=blue,urlcolor=blue]{hyperref}

\title{\textbf{State and spatial completion of 1D premixed flames for hydrogen and methane combustion}}

\author[1]{Federico~Rios~Tascon\thanks{Corresponding author. E-mail: \href{mailto:friostas@stanford.edu}{friostas@stanford.edu}}}
\author[2]{Ryan~F.~Johnson}
\author[1]{Diego~D.~Ortiz}
\author[3]{Peter~J.~Schmid}
\author[1]{Beverley~J.~McKeon}

\affil[1]{Stanford University, Stanford, CA 94305}
\affil[2]{U.S. Naval Research Laboratory, Washington, D.C. 20375}
\affil[3]{King Abdullah University of Science and Technology, Thuwal, Saudi Arabia}

\date{}

\begin{document}

\maketitle

\begin{abstract}
%%%
Accurate prediction of chemically reacting flows requires resolving the nonlinear coupling among transport and chemical reaction operators, often at substantial computational cost. This work investigates Fourier Neural Operators (FNOs) for learning mappings from partial spatial profiles to steady, one-dimensional, freely propagating hydrogen-air and methane-air flame solutions. Using a database of over 15,000 cases generated with Cantera, two different FNO models were trained to reconstruct the complete thermochemical state from a smaller set of input variables. The first model concerns state completion which uses only temperature, fuel, and oxygen mass-fraction profiles to reconstruct the remaining 7 species for hydrogen flames and 51 species for methane flames along with the missing velocity. The second model performs spatial completion, using upstream thermochemical profiles extending into the initial temperature rise to reconstruct the withheld downstream structure. Evaluation on 2,625 test cases per fuel, excluded from training, shows recovery of reaction-zone species profiles, including radicals and intermediates, across the sampled conditions.  These results demonstrate reconstruction of species and velocity profiles that are not supplied as inputs, and recovery of downstream structure outside the observed region. Reconstruction with missing species or incomplete spatial coverage suggests potential for using learned operators in estimating inaccessible thermochemical quantities and developing reduced-state representations for combustion applications.

%%%%
\end{abstract}

\noindent\textbf{Keywords:} Machine learning, Fourier Neural Operators, FNOs, Reduced order modeling

\section{Introduction}
The simulation of chemically reacting flows remains a major challenge in computational fluid dynamics (CFD) due to the stiffness and high dimensionality of the governing equations, the large number of chemical species and reactions involved, and the stringent mesh and temporal resolution requirements needed to resolve the thin reaction zones and steep gradients near flame fronts \citep{poinsot2005theoretical, oran2001numerical}. Reduced-order models (ROMs) have long been used to mitigate these costs and accelerate chemistry solvers \citep{lu2009toward}, but traditional approaches often rely on assumptions that limit their applicability beyond the regimes for which they were developed. 

Progress-variable approaches \citep{ihme2005prediction,Boj23,GONZALEZJUEZ201726,Clavin_Williams_1979,PETERS1984319,PITSCH19981057,PIERCE_MOIN_2004,KNUDSEN2009678} reduce state dimensionality by describing reacting flows with a small set of transported scalars that capture key combustion features. Canonical reacting-flow solutions are used to construct manifolds that map progress variables and other independent variables to the corresponding thermochemical states. These approaches compress the state representation, but their accuracy depends on how well the manifold represents the governing physics and conditions of interest \citep{vanOijen2000FGM}. In parallel, chemical mechanism reduction seeks to retain the effects of detailed kinetics while reducing the number of species and reactions \citep{lu2005directed,esposito2011skeletal}. Reduced mechanisms can contain roughly ten to thirty species, including some HyChem models for hydrocarbon fuels \citep{xu2018physics}. Despite these reductions, such mechanisms remain computationally demanding for large-scale simulations or iterative design studies. Both manifold-based approaches and reduced mechanisms can lose accuracy outside the thermochemical conditions for which they were developed and validated.

In addition, optical diagnostics such as OH planar laser-induced fluorescence (OH-PLIF) and infrared thermography enable noninvasive measurements of radical concentrations and temperature fields in reacting flows. However, these diagnostics do not directly provide the complete thermochemical state, leaving key quantities unmeasured or inferred indirectly. A reduced-order model (ROM) designed for state reconstruction could help bridge this gap by using limited observational inputs (for example, temperature and selected species signals) to estimate the remaining state variables. Such ROMs have the potential to enhance interpretation of OH-PLIF and thermographic data and to improve physical insight into combustion processes where direct measurement of all quantities is not practical or feasible \citep{Eckbreth1996,KohseHoinghaus2002,Alden2011}.

Recent advances in machine learning offer new opportunities to develop data-driven ROMs that generalize across conditions. Fourier Neural Operators (FNOs) \citep{li2021fourierneuraloperatorparametric} have shown strong capability in learning mappings between infinite-dimensional function spaces, motivating their use for combustion state prediction. Previous combustion studies have established the relevance of operator learning for combustion. Weng et al.\ \citep{weng2025extendedFNO} extended FNOs for stiff chemical kinetics and generalization to unseen conditions. Wu et al.\ \citep{wu2026resolutionInvariant} used a factorized FNO to reconstruct temperature and species fields from OH spatial data in auto-ignition-stabilized turbulent nonpremixed jet flames, including evaluations across resolutions.

This work examines whether a learned operator can recover the coupled structure of steady premixed flames from partial profiles across variations in reactant temperature, pressure, and equivalence ratio. Two complementary tasks are considered: reconstructing unprovided species and velocity from temperature and selected reactant profiles throughout the domain, and reconstructing downstream structure from the full thermochemical state in an upstream region. For the sampled hydrogen-air and methane-air flames, these tests demonstrate recovery of species profiles and downstream products from incomplete inputs. The operator maps spatial profiles to flame structure, differing from local lookup methods based on prescribed manifold coordinates.

\section{Dataset}\label{sec:dataset}

\begin{figure}[h!]
\centering
\includegraphics[width=0.6\textwidth]{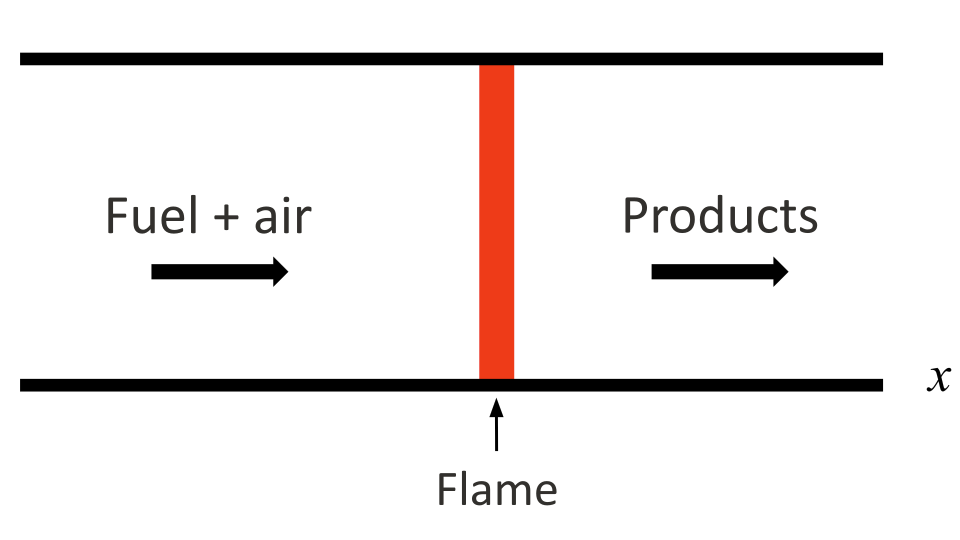}
\caption{Freely propagating premixed flame configuration.}
\label{flame_setup}
\end{figure}

This work considers a freely propagating premixed hydrogen-air flame, and a methane-air flame. In each case, hydrogen (H$_2$) or methane (CH$_4$) is premixed with oxygen (O$_2$) and nitrogen (N$_2$) at a specified temperature and pressure, forming a flame with a thin reaction zone where temperature and species concentrations have sharp features. Figure \ref{flame_setup} shows a diagram of this configuration.

For each fuel case, over 15,000 cases were generated for a range of specified reactant temperatures ($T$), pressures ($p$), and equivalence ratios ($\phi$). The ranges for these parameters are shown in Table \ref{tab:cantera_params} for the hydrogen-air flame, and in Table \ref{tab:cantera_params_methane} for the methane-air flame. For any choice of reactants $T$, $p$, and $\phi$, simulation provides the spatial distribution of temperature, velocity, and species mass fractions for a fixed domain length.

\begin{table}[ht!] \footnotesize
\caption{Ranges of temperatures ($T$), pressures ($p$), and equivalence ratios ($\phi$) used as parameters in Cantera for the hydrogen-air flame cases.}
\centerline{\begin{tabular}{lc}
\hline 
Parameter & Parameter range \\
        \hline
        $T$  & 300--800 K\\
        $p$ &  0.5--3 atm\\
        $\phi$  & 0.3--1.3\\
\hline 
\end{tabular}}
\label{tab:cantera_params}
\end{table}
\begin{table}[ht!] \footnotesize
\caption{Ranges of temperatures ($T$), pressures ($p$), and equivalence ratios ($\phi$) used as parameters in Cantera for the  methane-air flame cases.}
\centerline{\begin{tabular}{lc}
\hline 
Parameter & Parameter range \\
        \hline
        $T$  & 300--800 K\\
        $p$ &  0.5--3 atm\\
        $\phi$  & 0.5--1.3\\
\hline 
\end{tabular}}
\label{tab:cantera_params_methane}
\end{table}
The data for this flow was generated using Cantera \citep{goodwin2016cantera}, an open-source chemical kinetics software. In the hydrogen-air flame case, a chemical mechanism extracted from the Foundational Fuel Chemistry Model Version 2.0 (FFCM-2) \citep{ZDV2023} was implemented, which considers nine species, comprising 25 chemical reactions that describe hydrogen oxidation. The methane-air flame case used the GRI-Mech 3.0 chemical mechanism \citep{gri30}, with 53 chemical species considered for the combustion process, and a total of 325 chemical reactions.

\subsection{Flame recentering\label{subsec:recentering}} \addvspace{10pt}

Each of the flame solutions returned by Cantera has its own adaptive grid, so both the number of grid points and the streamwise position of the flame differ from case to case. The flame position of a given case is taken to be the location $x_{50\%}$ at which the temperature reaches the midpoint of its total rise,
\begin{align}
\label{eq:anchor}
    T(x_{50\%})=\tfrac{1}{2}\left(T_u+T_b\right),
\end{align}
where $T_u$ and $T_b$ denote the temperatures at the upstream and downstream boundaries of the computational domain. This flame location in the original data is dependent on the choice of $T$, $p$, and $\phi$. In order to have a consistent flame location between cases, each flame is shifted as $x\mapsto x-x_{50\%}+x_\mathrm{ref}$ onto a grid spanning $0\leq x\leq0.1$~m, common to every case, with the anchor placed at $x_\mathrm{ref}=0.03$~m. In the plots presented here, the location $x=0$~m represents the location of this anchor.
Flame alignment uses the reference-solution location $x_{50\%}$; the present tests therefore evaluate reconstruction in aligned coordinates rather than independent inference of flame position from partial observations.

\subsection{Data regridding\label{subsec:data_reg}} \addvspace{10pt}

As outlined in Section \ref{sec:FNO_arc}, each of the implemented FNO architectures utilizes the Fourier transform during the process of learning an operator on a set of the lowest modes. This Fourier transform component can struggle with high gradients in the data, where the Gibbs phenomenon might occur. To address this issue, as well as the problematic grid density, the one-dimensional flame front cases, generated by Cantera, were preprocessed before training the FNO. Knowledge of the different regions of the flame was used to regrid the data in order to add resolution to the average reaction zone location. 

\begin{figure}[h!]
\centering
\includegraphics[width=0.6\textwidth]{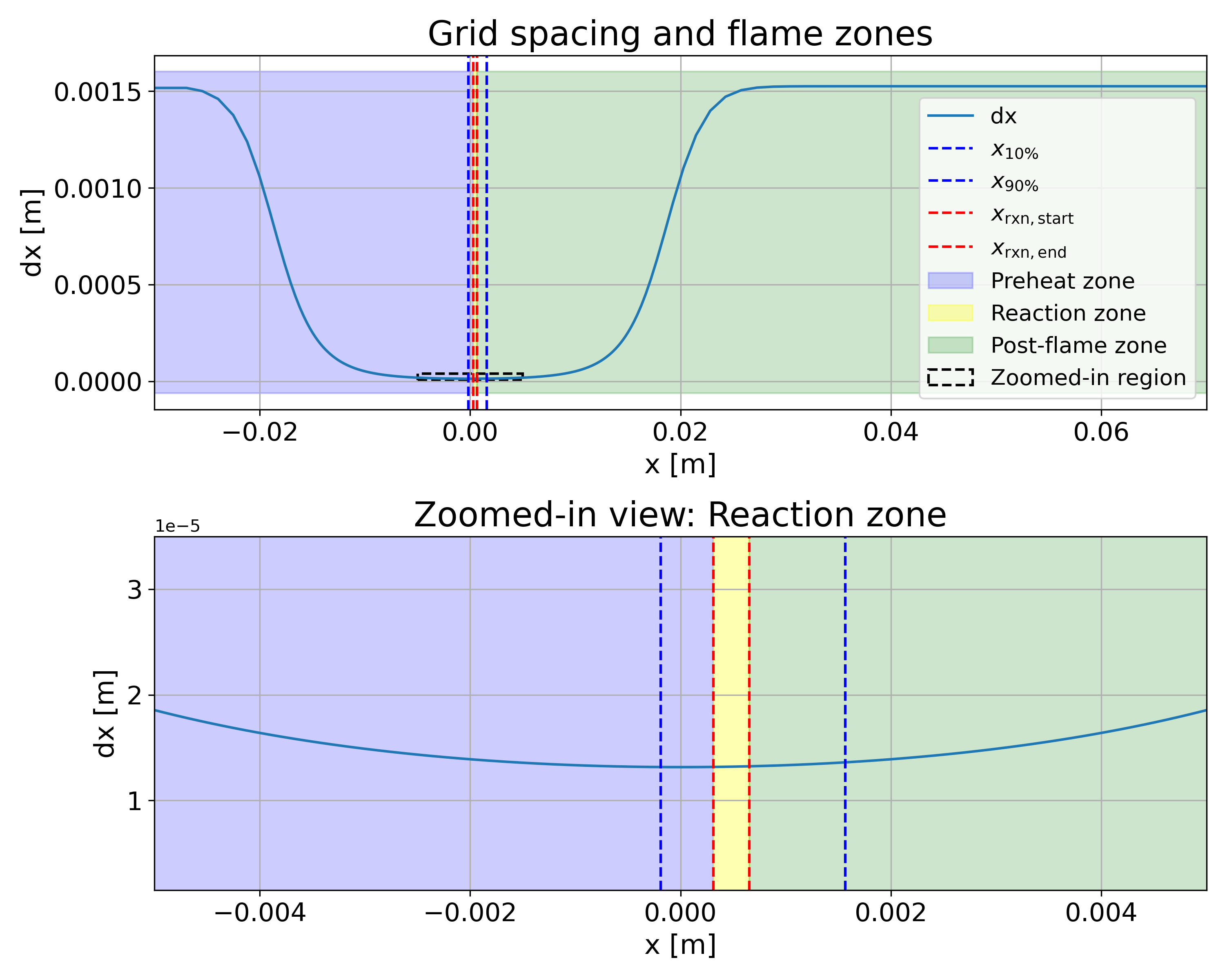}
\caption{Flame zones and grid spacing for a hydrogen premixed flame.}
\label{grid}
\end{figure}

Figure~\ref{grid} illustrates the standard dissection of a laminar premixed flame into three regions \citep{peters}. In the preheat zone, reactants are heated but no significant conversion has yet occurred. The reaction zone contains the primary heat release and species conversion, with steep temperature and mass-fraction gradients; this region sets the requirement for the smallest grid spacing. The locations where the temperature reaches 10\% and 90\% of its total rise, $x_{10\%}$ and $x_{90\%}$, indicate that most of the thermal transition occurs over a narrow interval near the reaction zone. Downstream, in the post-flame zone, reaction products are fully formed and gradients are weak. This structure indicates that the dominant thermochemical changes are confined to a very narrow region with a denser mesh than in the rest of the domain. This characteristic increases the difficulty of learning pertinent features during training.

The grid spacing ($\Delta x$) used for the hydrogen flame case, interpolated from its original adaptive Cantera grid onto a new non-uniform grid, is also shown in Figure \ref{grid}. The new grid was generated using an inverse cumulative distribution function (CDF) approach, where a spatial point density function is first defined as a Gaussian-augmented uniform distribution, as shown in Equation \ref{eq:CDF_grid}. 
\begin{align}
\label{eq:CDF_grid}
    \rho(x)=1+A\exp\left(-\frac{(x-x_f)^2}{2\sigma^2}\right)
\end{align}
In this equation, $x_f$ is the mean flame location, $\sigma$ controls the width of the refinement region, and $A$ sets the amplitude of the density enhancement relative to the background uniform distribution. The parameters in Equation \ref{eq:CDF_grid} were set to $x_f=0.03$ m, $\sigma = 0.006$ m, and $A=115$ for both the hydrogen and methane flame cases. It is important to note that this data regridding is not unique or strictly necessary. Any type of grid refinement around the flame region would help avoid possible Gibbs-phenomenon artifacts, but as long as the implemented FNO contains sufficient trainable parameters, even data with a uniform grid will perform well.

\section{FNO architectures\label{sec:FNO_arc}}
Two different FNO architectures were developed to take a reduced set of input state variables and output the full thermochemical state, including sharp species profiles in aligned coordinates. In each case, $10,000$ flow cases were used as training data and $3,000$ cases as validation data. An additional $2,625$ cases were left as test data.

Throughout this and following sections, variables with a hat ($\wedge$) represent outputs of the FNOs. For all FNOs, the relative $L^2$-error was used as the loss function, as defined by Equation \ref{eq:loss}.
\begin{align}
    \label{eq:loss}
    \text{Relative }L^2\text{-error}=\frac{||Y-\hat{Y}||_2}{||Y||_2}
\end{align}
This loss measures the discrepancy between predicted and reference profiles across the entire spatial domain, ensuring that the features in every region of the flame are captured during optimization.

Each of the FNOs described in this section, shown in Figures \ref{state_FNO} and \ref{spatial_FNO}, comprises three main components: an encoder, a decoder, and a single linear Fourier layer. The encoder ($P$) maps the input fields into a higher-dimensional latent representation using two one-dimensional convolutional layers followed by Gaussian error linear unit (GELU) activation functions. Inside the Fourier layer, the top section learns an operator ($R$) for a set of the lowest Fourier modes, while the bottom section ($W$) is a single linear convolutional layer that acts as a bias. Finally, the decoder ($Q$) maps the latent representation back to the physical variables using two additional one-dimensional convolutional layers with a GELU activation between them \citep{li2021fourierneuraloperatorparametric}. A single linear Fourier layer in each FNO helps avoid unnecessarily deep networks while also reducing the risk of overfitting, and also enables the use of the FNO as an input-output model to learn functional relationships between different sub-processes in the data. One of the design choices in this case was the size of the encoder, decoder, and Fourier layer. For both architectures and fuel cases, each of the convolutional layers in the encoder and decoders was set to have $96$ neurons, while $32$ Fourier modes were kept inside the Fourier layer. The parameter optimization for all models during training was performed using the Adaptive Moment Estimation (Adam) algorithm \citep{kingma2017adammethodstochasticoptimization} with a learning rate set to $1\times10^{-3}$. In each case, the training procedure was performed for $60$ epochs. The parameter choices for all implemented architectures are summarized in Table \ref{tab:repro_table}.

\begin{table}[h!]
\caption{Parameters and architecture implemented in all FNO models considered in this work.}
\centering\footnotesize
\begin{tabular}{@{}l@{\hspace{1em}}>{\raggedright\arraybackslash}p{0.53\columnwidth}@{}}
\hline
\textbf{Item} & \textbf{Value} \\
\hline
\multicolumn{2}{@{}l}{\textbf{Architectures}} \\
Encoder $P$ & 2 convolutional layers, GELU after each \\
Fourier layer & single and strictly linear \\
Decoder $Q$ & 2 convolutional layers, GELU between them \\
Width & 96 \\
Fourier modes & 32 \\
Branch combination & $0.7\,R(x) + 0.3\,W(x)$ \\
Fourier weight init. & $0.02\times\mathcal{N}(0,1)$, complex \\
Conv.\ weight init. & PyTorch default (Kaiming uniform) \\
Internal normalizer & channel-wise $z$-score, fitted on the first training batch \\
\hline
\multicolumn{2}{@{}l}{\textbf{Training}} \\
Loss & relative $L^2$ error, averaged over the batch \\
Optimizer & Adam \\
Learning rate & $1\times10^{-3}$ \\
Batch size & 16 \\
Epochs & 60 \\
Gradient clipping & none \\
\hline
\multicolumn{2}{@{}l}{\textbf{Implementation}} \\
Framework & PyTorch \citep{paszke2019pytorch} \\
Random seeds & \texttt{torch.manual\_seed(0)},
               \texttt{np.random.seed(0)} \\
Precision & float32 \\
Device & single CUDA GPU \\
\hline
\end{tabular}
\label{tab:repro_table}
\end{table}

\subsection{State-completing FNO\label{subsec:state_comp_FNO}}
\begin{figure}[!t]
\centering
\includegraphics[width=0.75\textwidth]{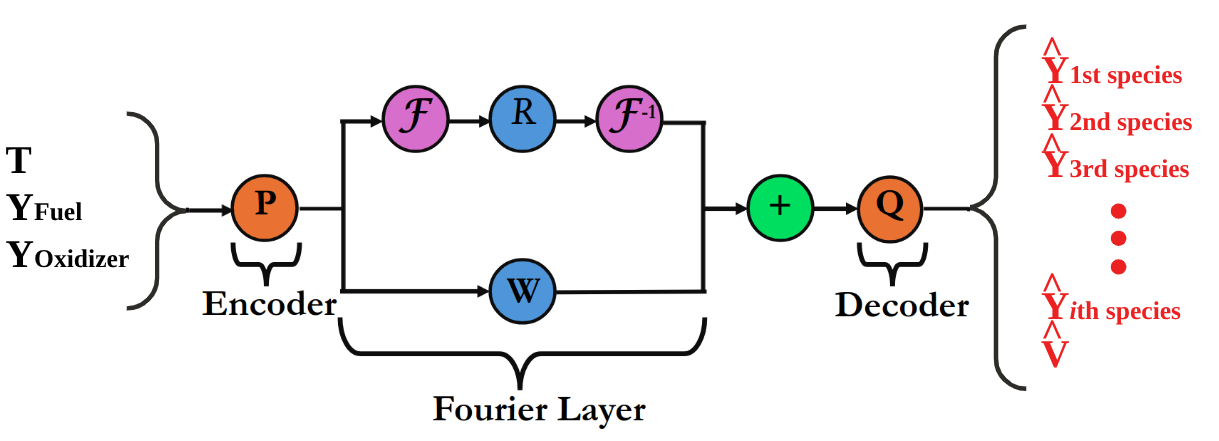}
\caption{State-completing FNO architecture.}
\label{state_FNO}
\end{figure}

The first architecture, shown in Figure \ref{state_FNO}, completes the thermochemical and fluid-dynamic state of the premixed flame across the full spatial domain using only temperature and the fuel and oxidizer mass fractions as inputs, yielding the other species mass fractions and velocity field. Within the sampled flame conditions, these three input profiles support reconstruction of the remaining species and velocity field across variations in reactant temperature, pressure, and equivalence ratio.

\subsection{Spatially-completing FNO\label{subsec:spatially_comp_FNO}}

\begin{figure}[!t]
\centering
\includegraphics[width=\textwidth]{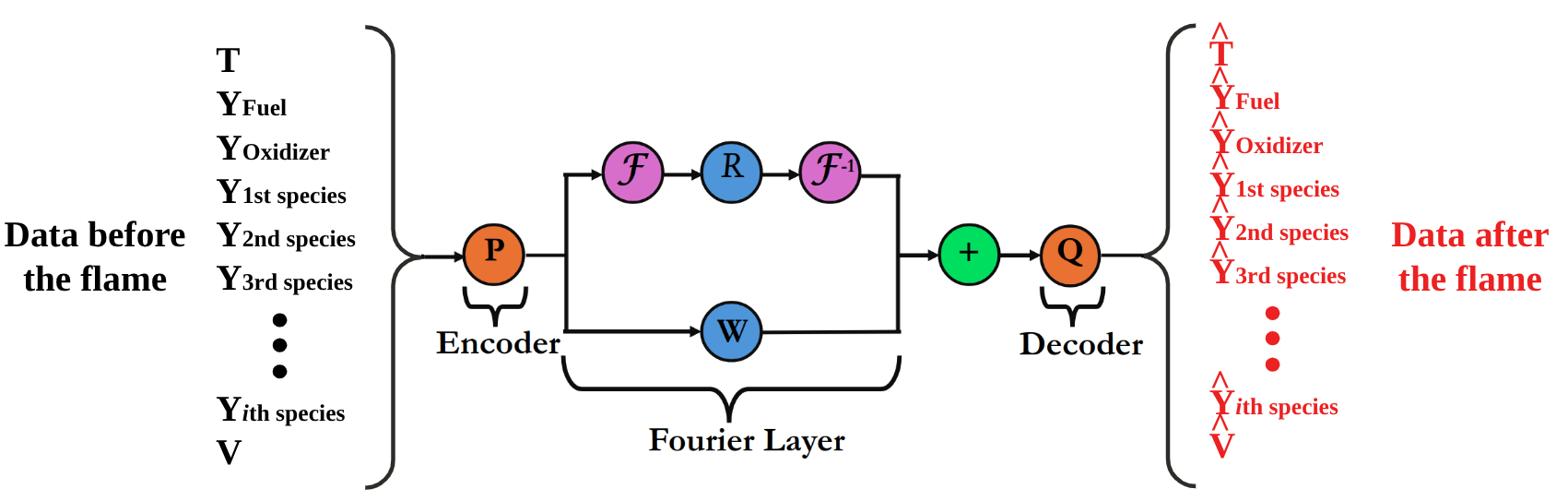}
\caption{Spatially-completing FNO architecture.}
\label{spatial_FNO}
\end{figure}

Figure \ref{spatial_FNO} shows the second implemented FNO architecture. In this case, the FNO takes the full thermochemical state of the flow before $x_{10\%}$ (the average location where the temperature reaches $10\%$ of its total rise), and outputs the thermochemical state for the spatial region after this location. The pressure information for this input is implicit in the shape of the curves between $x_{0\%}$ and $x_{10\%}$ (the spatial region between locations where the temperature starts to reach up to $10\%$ of its total rise). The profiles of the thermochemical state, for different pressures but the same reactant temperature and equivalence ratios, will look the same for the region before the temperature starts to rise ($x_{0\%}$). Consequently, a small part of the region, where temperature starts to increase, must be included in the input.

\section{Results\label{sec:FNO_res}} 

The prediction error for each species mass fraction was quantified using the weighted mean squared error, $E_W$, where the weights correspond to the grid spacing ($\Delta x_i$). The $E_W$ is defined as
\begin{align}
\label{wmse}
    E_{W}=\frac{1}{n}\frac{\sum_{i=1}^n \Delta x_i(Y_{i}-\hat{Y}_{i})^2}{\sum_{i=1}^n \Delta x_i}
\end{align}
where $Y_{i}$ and $\hat{Y}_{i}$ denote the true and predicted mass fractions, and $n$ represents the number of grid points. To complement $E_W$, the maximum pointwise error for each species was computed as
\begin{align}
\label{maxError}
    E_{\max} = \max_i |Y_{i} - \hat{Y}_{i}|\text{.}
\end{align}
The variances of both metrics across the test set were computed using
\begin{align}
\label{var}
    &\mathrm{Var}(E_W)=\frac{1}{m}\sum_{j=1}^m(E_{W,j}-\overline{E}_{W})^2\quad \text{ and }\\
    \label{varMax}
    &\mathrm{Var}(E_{\max})=\frac{1}{m}\sum_{j=1}^m(E_{\max,j}-\overline{E}_{\max})^2.
\end{align}
where $m$ is the number of cases in the test set. These quantities are both global and localized performance error metrics. The weighted mean squared error emphasizes accuracy across the entire spatial domain while accounting for the nonuniform grid spacing introduced during preprocessing. In contrast, the maximum pointwise error highlights the worst-case discrepancy between predicted and reference values. 

\subsection{State-completing FNO results for a hydrogen flame\label{subsec:state_comp_FNO_res}} 
\begin{figure}[h!]
\centering
\includegraphics[width=0.6\textwidth]{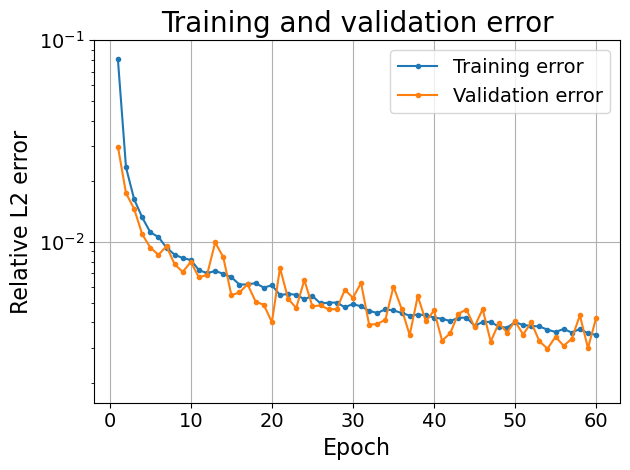}
\caption{Training errors for state completion of a hydrogen-air flame measured using Equation \ref{eq:loss}.}
\label{train_err_hyd}
\end{figure}

This section presents the results obtained by implementing the FNO architecture described in Section \ref{subsec:state_comp_FNO} for the hydrogen flame data. 
Figure \ref{train_err_hyd} shows the training and validation errors, calculated using Equation \ref{eq:loss}, during the training procedure. This figure shows good convergence of the model weights without any signs of overfitting, with both the training and validation errors having similar magnitudes after $60$ epochs.

\begin{figure}[h!]
\centering
\includegraphics[width=0.6\textwidth]{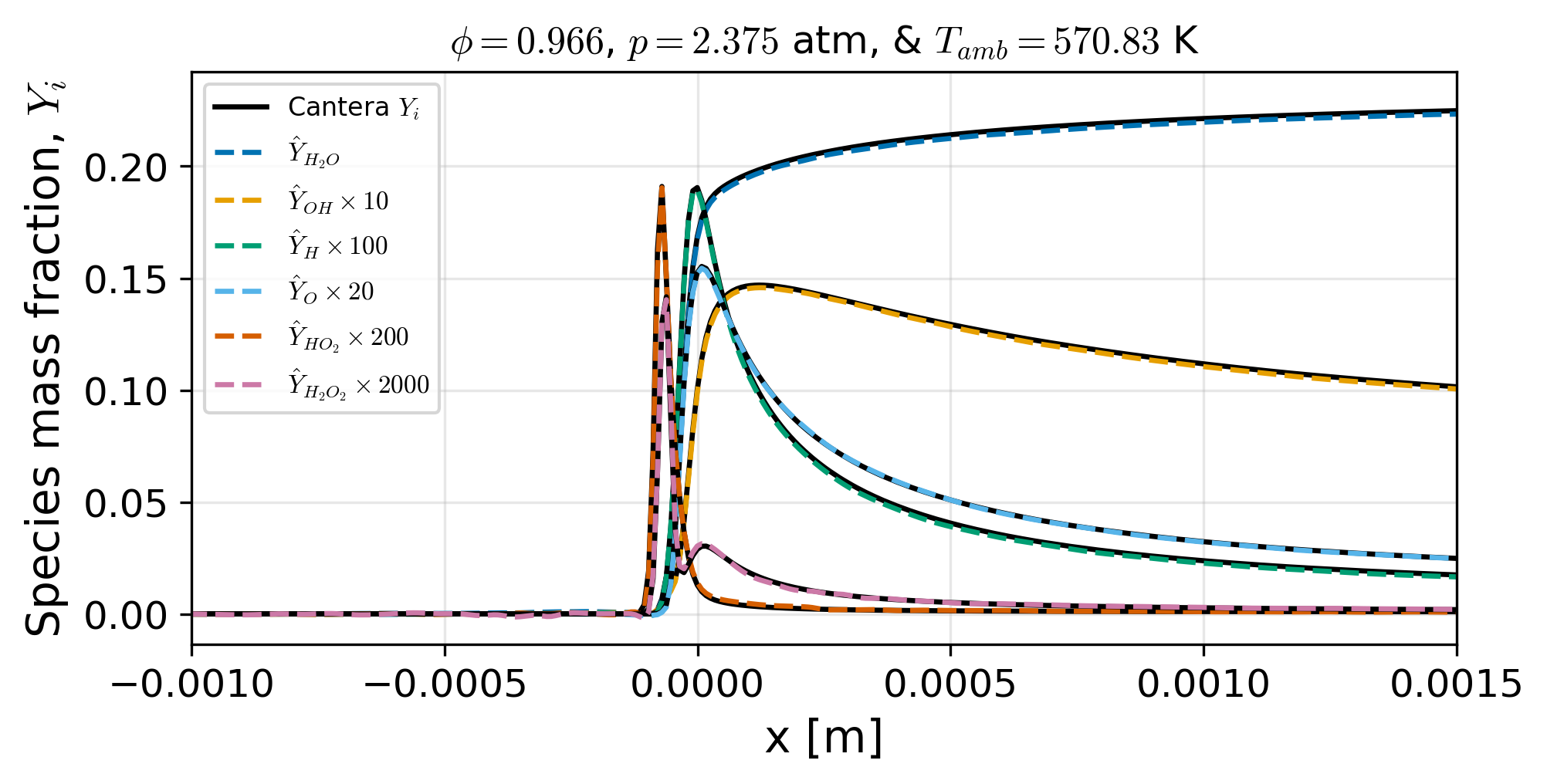}
\caption{State-completing FNO mass fraction predictions for a single representative test case of a hydrogen flame. The $x$-axis origin is the recentering anchor $x_{50\%}$, the location where the temperature reaches $50\%$ of its total rise.}
\label{FNO_state_pred}
\end{figure}

Figure~\ref{FNO_state_pred} shows premixed flame solutions, where each line color corresponds to a species mass fraction for $Y_{\text{H}_2\text{O}}$, $Y_{\text{OH}}$, $Y_{\text{H}}$, $Y_{\text{O}}$, $Y_{\text{H}\text{O}_2}$, and $Y_{\text{H}_2\text{O}_2}$. Solid black lines denote the Cantera solutions, and dashed lines denote the FNO predictions. The presented case uses a sample within the bounds of but explicitly \emph{not included in the training or validation sets}: $\phi = 0.966$, $p = 2.375$~atm, $T_{amb} = 570.83$~K. The $x$-axis is centered around the $x_{50\%}$ location for this particular flow configuration.

The metrics defined in Equations \ref{wmse}, \ref{maxError}, \ref{var}, and \ref{varMax} were computed for $2,625$ test cases, i.e., $m=2,625$, and are listed in Table~\ref{table:wmse} and Table~\ref{table:maxerror} for $E_W$ and $E_{\max}$, respectively. These results demonstrate that the FNO model accurately captures the spatial structure of the premixed flame. In particular, the model successfully predicts the sharp gradients associated with the reaction zone, as well as the smoother profiles in the preheat and post-flame regions. The small variances of both $E_W$ and $E_{\max}$ indicate that this performance is consistent across the full range of thermodynamic conditions represented in the test dataset. The largest errors occur for H$_2$O and for N$_2$ which are the species with the highest mass fractions in the post-flame region. These larger absolute errors occur because even small relative deviations correspond to larger pointwise differences when the baseline mass fractions are large.

\begin{table}[h!]
\begin{center}\scriptsize
\vskip -0.05in
\begin{tabular}{c c c}
\textbf{Variable} & \textbf{$E_W$} & \textbf{$\mathrm{Var}(E_W)$} \\
\hline
$Y_{\text{H}_2\text{O}}$       & $6.12\times10^{-10}$ & $2.37\times10^{-19}$ \\
$Y_{\text{O}\text{H}}$         & $1.74\times10^{-12}$ & $3.91\times10^{-24}$ \\
$Y_{\text{N}_2}$               & $9.75\times10^{-11}$ & $1.48\times10^{-20}$ \\
$Y_{\text{H}}$                 & $2.78\times10^{-14}$ & $2.55\times10^{-27}$ \\
$Y_{\text{O}}$                 & $1.88\times10^{-13}$ & $6.11\times10^{-26}$ \\
$Y_{\text{H}\text{O}_2}$       & $2.04\times10^{-15}$ & $3.57\times10^{-30}$ \\
$Y_{\text{H}_2\text{O}_2}$     & $1.20\times10^{-16}$ & $9.93\times10^{-32}$ \\
$V$                            & $4.71\times10^{-5}$ & $3.42\times10^{-9}$ \\
\hline
\end{tabular}
\caption{Average weighted mean square error ($E_W$) and its variance for each species mass fraction and the velocity predicted by the state-completing FNO across $2,625$ hydrogen-air test flow cases. The averages and variances were computed using Equations~\ref{wmse} and~\ref{var}. Values for $V$ are in $\mathrm{m}^2\,\mathrm{s}^{-2}$; all others are dimensionless.}
\label{table:wmse}
\end{center}
\end{table}

\begin{table}[h!]
\begin{center}\scriptsize
\vskip -0.05in
\begin{tabular}{c c c}
\textbf{Variable} & \textbf{$E_{\max}$} & \textbf{$\mathrm{Var}(E_{\max})$} \\
\hline
$Y_{\text{H}_2\text{O}}$       & $1.55\times10^{-3}$ & $4.24\times10^{-7}$ \\
$Y_{\text{O}\text{H}}$         & $1.41\times10^{-4}$ & $5.39\times10^{-9}$ \\
$Y_{\text{N}_2}$               & $1.02\times10^{-3}$ & $1.63\times10^{-7}$ \\
$Y_{\text{H}}$                 & $1.90\times10^{-5}$ & $1.23\times10^{-10}$ \\
$Y_{\text{O}}$                 & $7.76\times10^{-5}$ & $1.18\times10^{-9}$ \\
$Y_{\text{H}\text{O}_2}$       & $2.14\times10^{-5}$ & $2.61\times10^{-10}$ \\
$Y_{\text{H}_2\text{O}_2}$     & $3.44\times10^{-6}$ & $9.33\times10^{-12}$ \\
$V$                            & $3.25\times10^{-1}$ & $1.79\times10^{-2}$ \\
\hline
\end{tabular}
\caption{Average maximum pointwise error ($E_{\max}$) and its variance for each species mass fraction and the velocity predicted by the state-completing FNO across $2,625$ hydrogen-air test flow cases. The averages and variances were computed using Equations~\ref{maxError} and~\ref{varMax}. Values for $V$ are in $\mathrm{m}\,\mathrm{s}^{-1}$; all others are dimensionless.}
\label{table:maxerror}
\end{center}
\end{table}

\begin{figure}[h!]
\centering
\includegraphics[width=0.6\textwidth]{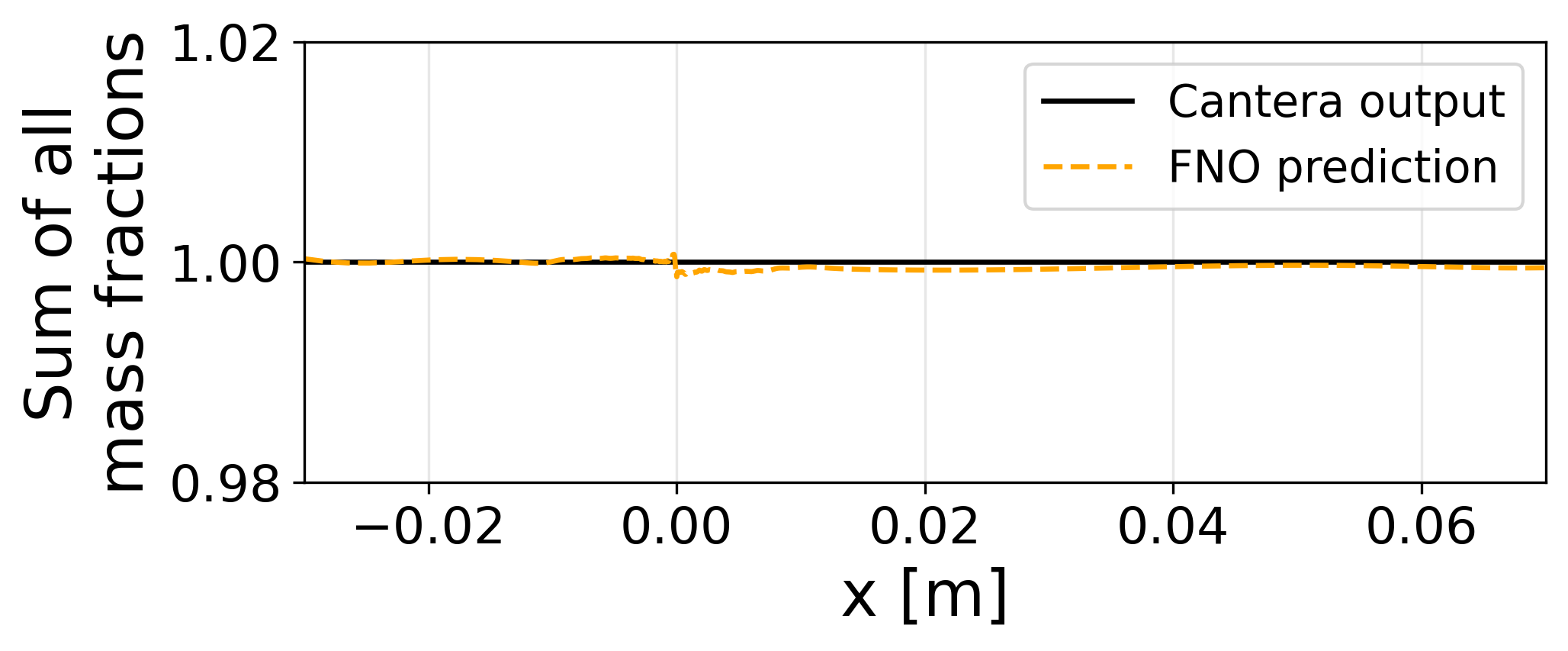}
\caption{Average sum of mass fractions over $2,625$ test flow cases for the hydrogen flame configuration. The $x$-axis origin is the recentering anchor $x_{50\%}$, the location where the temperature reaches $50\%$ of its total rise.}
\label{FNO_sumY}
\end{figure}

As an additional test of performance, the sum of all mass fractions (both those used as inputs and FNO outputs), was computed using Equation \ref{sum_massFrac}.
\begin{align}
\label{sum_massFrac}
    \text{Sum of mass fractions}=\sum_iY_i+\sum_j\hat{Y}_j
\end{align}
Due to the definition of mass fractions, a physically consistent prediction from the FNO requires that Equation \ref{sum_massFrac} equals one at every point in space. Figure \ref{FNO_sumY} shows the average of this sum over the $2,625$ test flow cases. Even though Equation \ref{sum_massFrac} is not explicitly enforced in the FNO, this figure shows that, on average, the FNO predictions remain very close to this constraint.

\subsection{State-completing FNO results for a methane flame\label{subsec:state_comp_FNO_res_methane}} 

\begin{figure}[h!]
\centering
\includegraphics[width=0.6\textwidth]{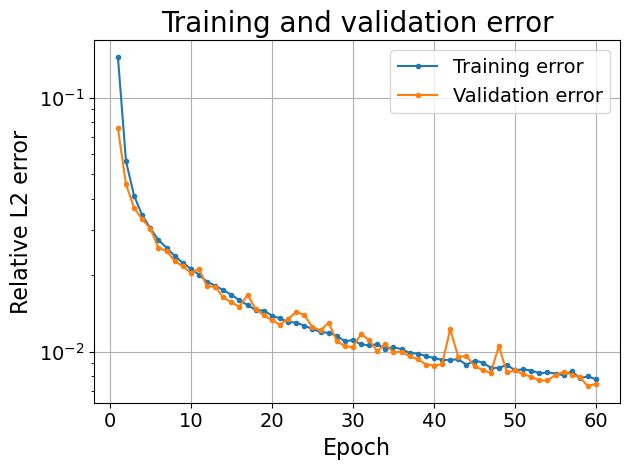}
\caption{Training errors for state completion of a methane-air flame measured using Equation \ref{eq:loss}.}
\label{train_err_methane}
\end{figure}

This section presents the results of the state-completing FNO architecture applied to the methane-air flame case, where the temperature ($T$), and the mass fractions for methane and oxygen ($Y_{\text{CH}_4}$ and $Y_{\text{O}_2}$) are used as inputs into the model to reconstruct the remaining $51$ species and velocity ($Y_{\text{H}_2}$, $Y_{\text{H}}$, $Y_{\text{O}}$, $Y_{\text{OH}}$, $Y_{\text{H}_2\text{O}}$, $Y_{\text{HO}_2}$, $Y_{\text{H}_2\text{O}_2}$, $Y_{\text{C}}$, $Y_{\text{CH}}$, $Y_{\text{CH}_2}$, $Y_{\text{CH}_2\text{(S)}}$, $Y_{\text{CH}_3}$, $Y_{\text{CO}}$, $Y_{\text{CO}_2}$, $Y_{\text{HCO}}$, $Y_{\text{CH}_2\text{O}}$, $Y_{\text{CH}_2\text{OH}}$, $Y_{\text{CH}_3\text{O}}$, $Y_{\text{CH}_3\text{OH}}$, $Y_{\text{C}_2\text{H}}$, $Y_{\text{C}_2\text{H}_2}$, $Y_{\text{C}_2\text{H}_3}$, $Y_{\text{C}_2\text{H}_4}$, $Y_{\text{C}_2\text{H}_5}$, $Y_{\text{C}_2\text{H}_6}$, $Y_{\text{HCCO}}$, $Y_{\text{CH}_2\text{CO}}$, $Y_{\text{HCCOH}}$, $Y_{\text{N}}$, $Y_{\text{NH}}$, $Y_{\text{NH}_2}$, $Y_{\text{NH}_3}$, $Y_{\text{NNH}}$, $Y_{\text{NO}}$, $Y_{\text{NO}_2}$, $Y_{\text{N}_2\text{O}}$, $Y_{\text{HNO}}$, $Y_{\text{CN}}$, $Y_{\text{HCN}}$, $Y_{\text{H}_2\text{CN}}$, $Y_{\text{HCNN}}$, $Y_{\text{HCNO}}$, $Y_{\text{HOCN}}$, $Y_{\text{HNCO}}$, $Y_{\text{NCO}}$, $Y_{\text{N}_2}$, $Y_{\text{AR}}$, $Y_{\text{C}_3\text{H}_7}$, $Y_{\text{C}_3\text{H}_8}$, $Y_{\text{CH}_2\text{CHO}}$, $Y_{\text{CH}_3\text{CHO}}$, $V$). Figure \ref{train_err_methane} shows the convergence of the training and validation errors during the training process of the network. As in the hydrogen case, this plot shows good convergence of the model's weights during the optimization procedure, with no obvious signs of overfitting.

\begin{figure}[h!]
\centering
\includegraphics[width=0.6\textwidth]{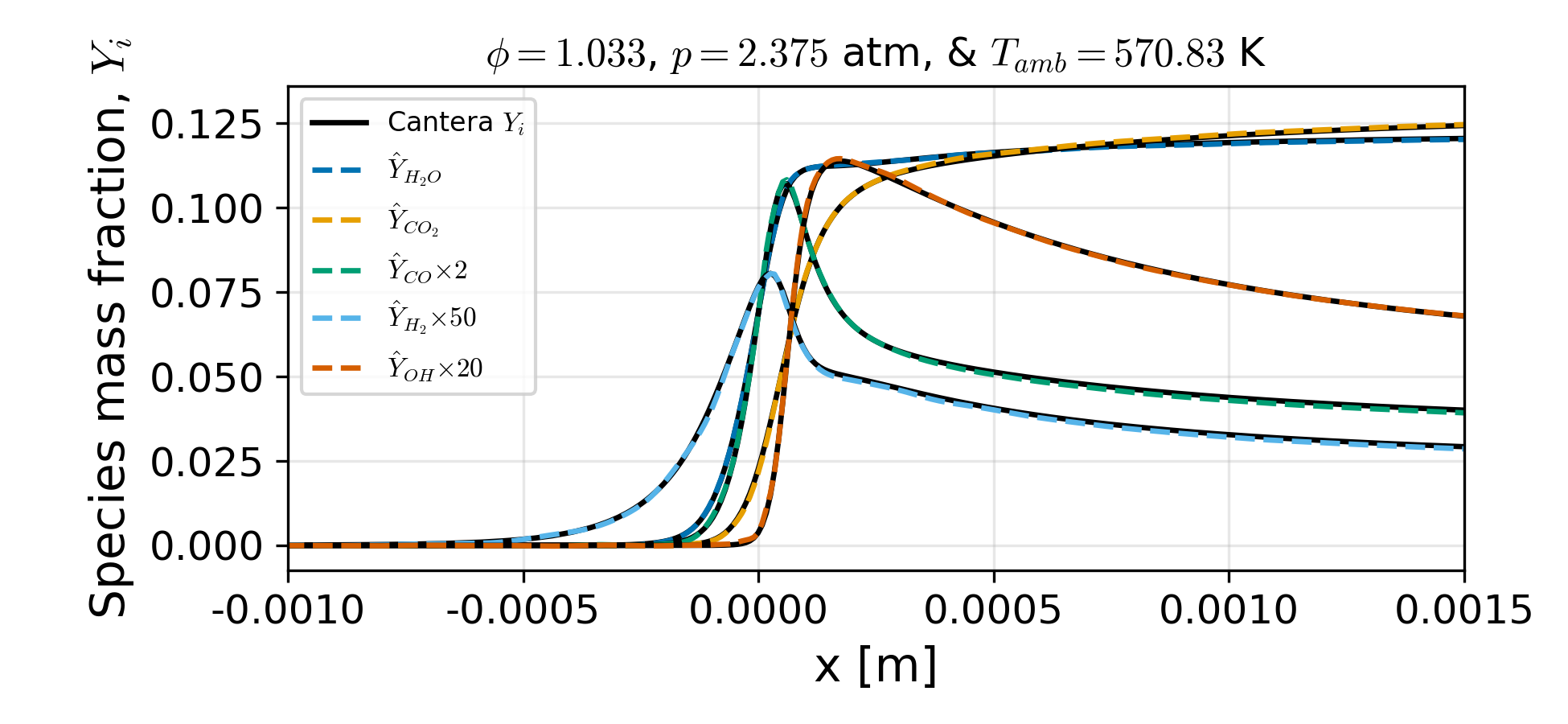}
\caption{State-completing FNO mass fraction predictions for a set of major combustion species for a single representative test case of a methane flame. The $x$-axis origin is the recentering anchor $x_{50\%}$, the location where the temperature reaches $50\%$ of its total rise.}
\label{FNO_state_pred_ch4_major}
\end{figure}

\begin{figure}[h!]
\centering
\includegraphics[width=0.6\textwidth]{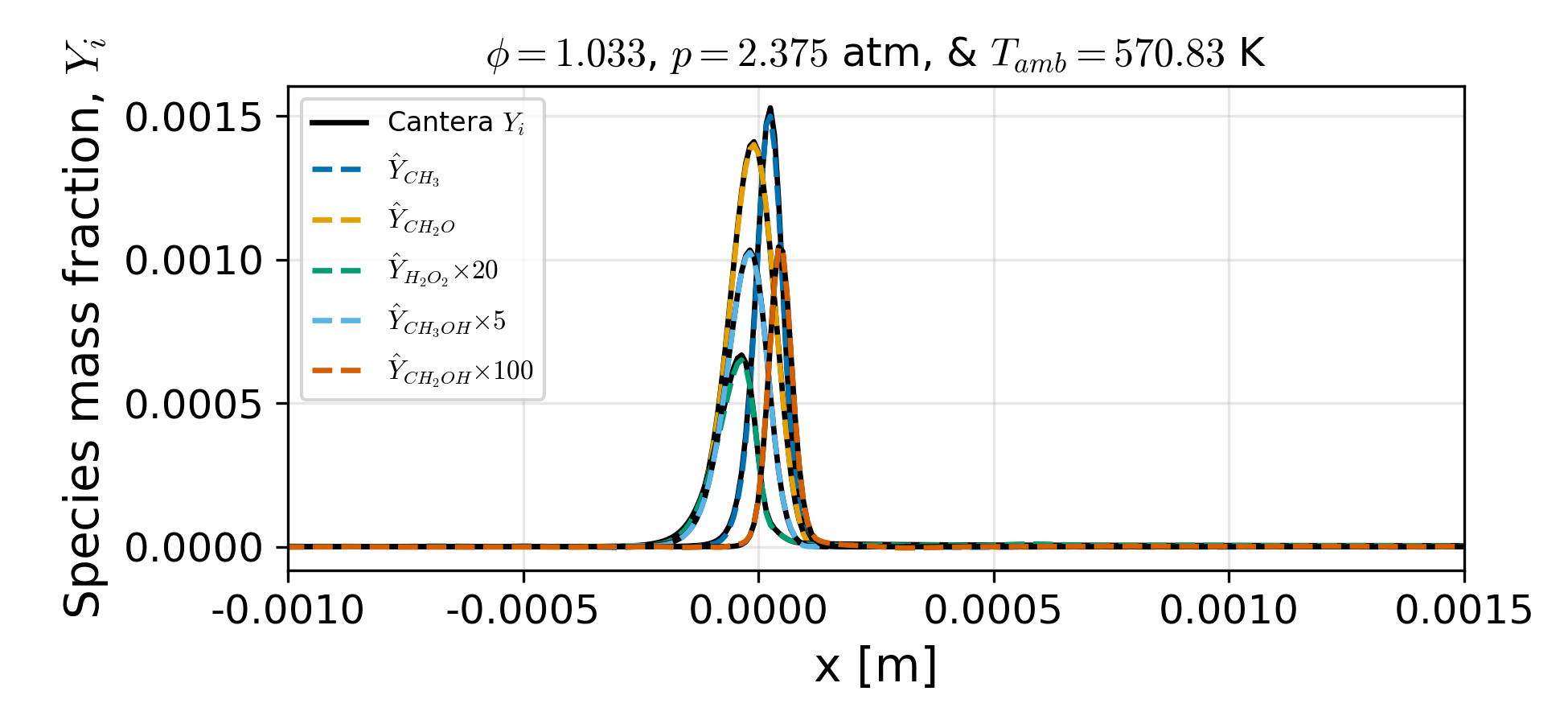}
\caption{State-completing FNO mass fraction predictions for a set of intermediate combustion species for a single representative test case of a methane flame. The $x$-axis origin is the recentering anchor $x_{50\%}$, the location where the temperature reaches $50\%$ of its total rise.}
\label{FNO_state_pred_ch4_int1}
\end{figure}

\begin{figure}[h!]
\centering
\includegraphics[width=0.6\textwidth]{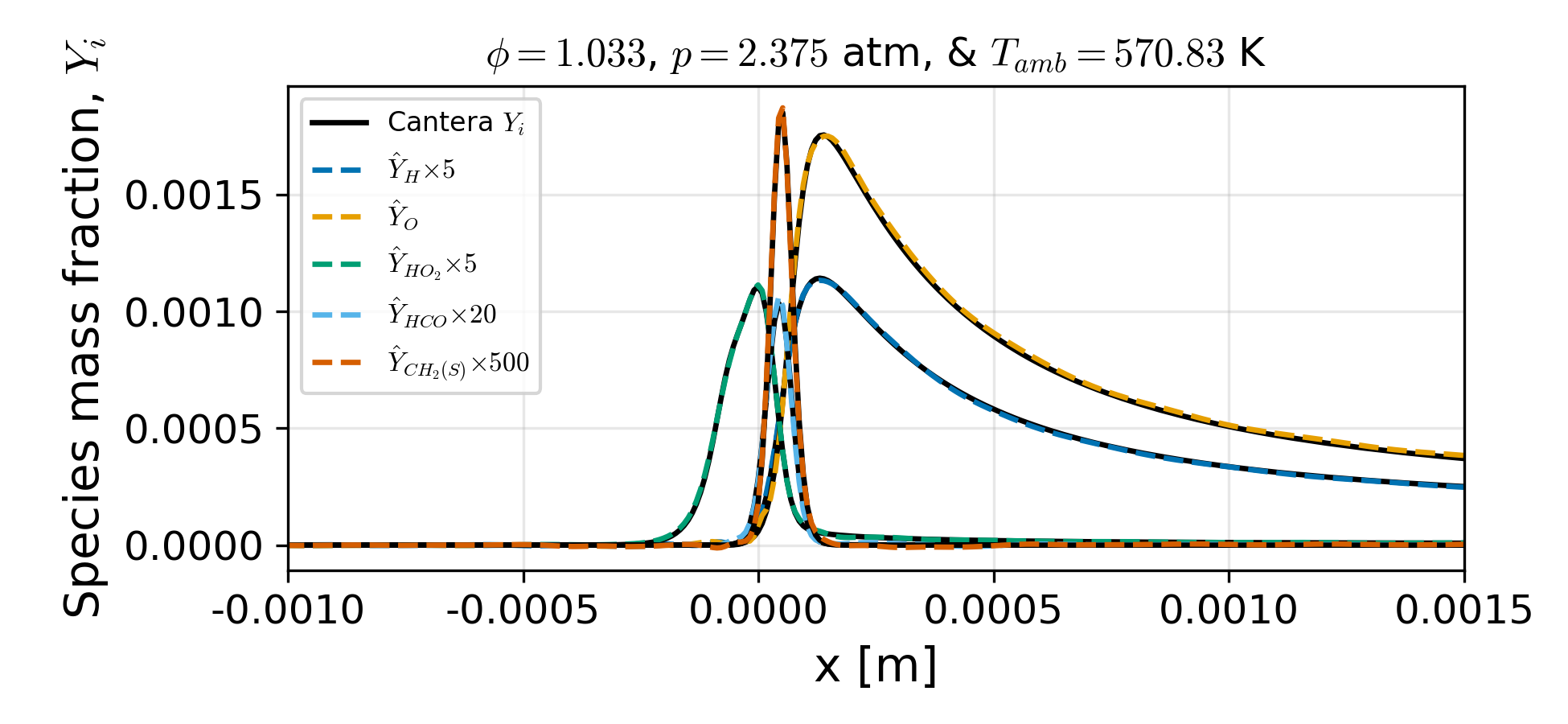}
\caption{State-completing FNO mass fraction predictions for a set of additional intermediate combustion species for a single representative test case of a methane flame. The $x$-axis origin is the recentering anchor $x_{50\%}$, the location where the temperature reaches $50\%$ of its total rise.}
\label{FNO_state_pred_ch4_int2}
\end{figure}

\begin{figure}[h!]
\centering
\includegraphics[width=0.6\textwidth]{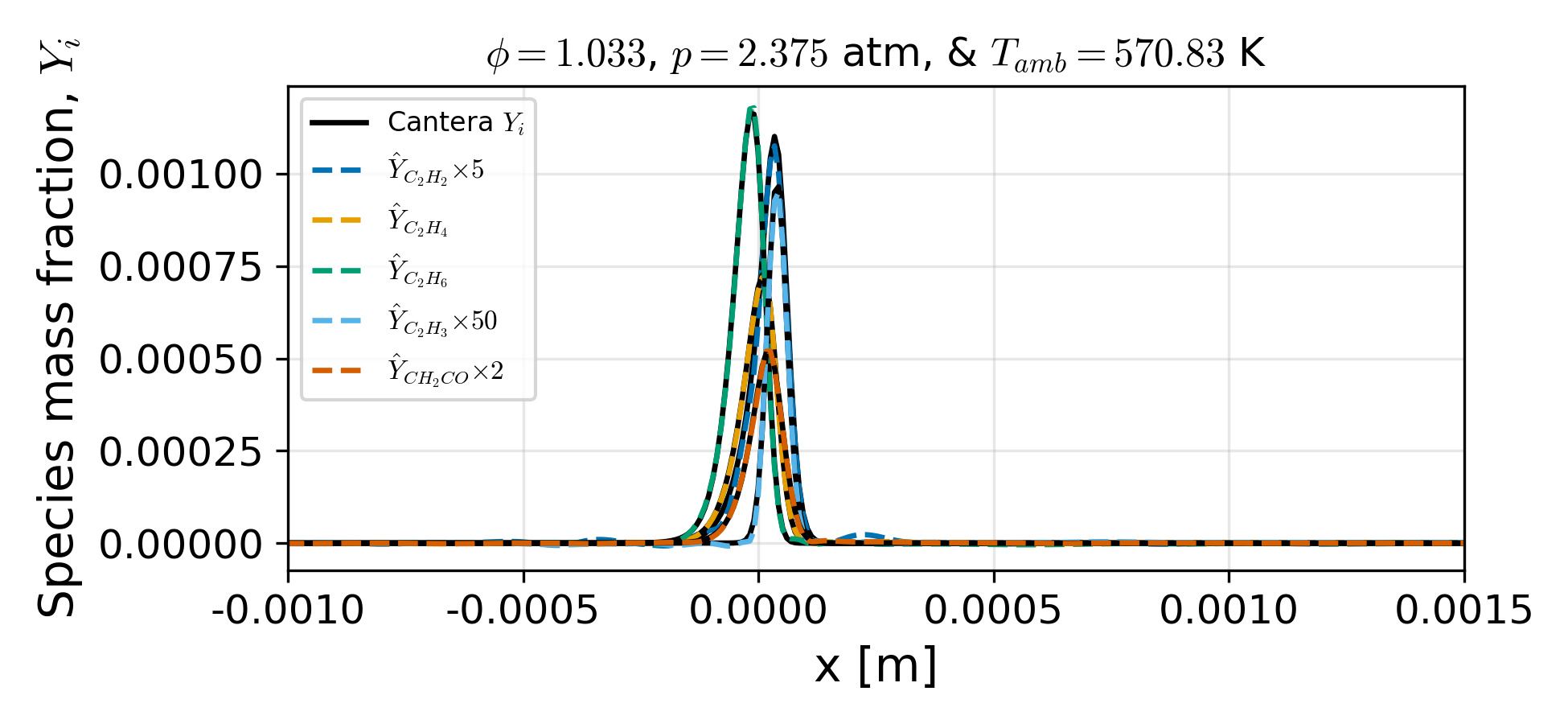}
\caption{State-completing FNO mass fraction predictions for a set of trace combustion species for a single representative test case of a methane flame. The $x$-axis origin is the recentering anchor $x_{50\%}$, the location where the temperature reaches $50\%$ of its total rise.}
\label{FNO_state_pred_ch4_trace}
\end{figure}

Figures~\ref{FNO_state_pred_ch4_major}--\ref{FNO_state_pred_ch4_trace} show the FNO predictions for a single representative case for a select set of species that are present in major, intermediate, and trace amounts during the combustion process. The presented case was generated with physical parameters $\phi = 1.033$, $p = 2.375$~atm, and $T_{amb} = 570.83$~K. In each figure, solid black lines denote the Cantera reference solution and dashed lines denote the FNO prediction. Across all species groups, the FNO is able to accurately predict the sharp gradients arising in the reaction zone.

Tables~\ref{table:wmse_methane} and~\ref{table:maxerror_methane} report the $E_W$ and $E_{\max}$ errors, along with their variances, for the species plotted in Figures \ref{FNO_state_pred_ch4_major}--\ref{FNO_state_pred_ch4_trace} computed over $2,625$ test cases. As in the hydrogen case, the largest absolute errors, both in $E_W$ and $E_{\max}$, occur for the species with the largest mass fractions in the post-flame region. This trend is consistent with the definitions of $E_W$ and $E_{\max}$ as absolute errors instead of relative errors, since because there is no normalization of the errors, the species with the larger mass fractions will produce a larger absolute error. However, it is worth noting that the magnitude of these errors is small compared to the mass fractions themselves, showing that there is good agreement, both globally and locally, between the predicted and reference profiles.

\begin{table}[h!]
\begin{center}\scriptsize
\vskip -0.05in
\begin{tabular}{c c c}
\textbf{Variable} & \textbf{$E_W$} & \textbf{$\mathrm{Var}(E_W)$} \\
\hline
$Y_{\text{H}_2\text{O}}$       & $1.97\times10^{-10}$ & $4.17\times10^{-20}$ \\
$Y_{\text{C}\text{O}_2}$       & $4.81\times10^{-10}$ & $2.14\times10^{-19}$ \\
$Y_{\text{C}\text{O}}$         & $6.67\times10^{-11}$ & $6.57\times10^{-21}$ \\
$Y_{\text{H}_2}$               & $2.13\times10^{-13}$ & $1.88\times10^{-25}$ \\
$Y_{\text{O}\text{H}}$         & $1.31\times10^{-12}$ & $2.55\times10^{-24}$ \\
$Y_{\text{C}\text{H}_3}$       & $2.99\times10^{-14}$ & $2.01\times10^{-27}$ \\
$Y_{\text{C}\text{H}_2\text{O}}$ & $1.06\times10^{-14}$ & $2.21\times10^{-28}$ \\
$Y_{\text{H}_2\text{O}_2}$     & $2.73\times10^{-17}$ & $2.02\times10^{-33}$ \\
$Y_{\text{C}\text{H}_3\text{O}\text{H}}$ & $1.91\times10^{-16}$ & $7.58\times10^{-32}$ \\
$Y_{\text{C}\text{H}_2\text{O}\text{H}}$ & $8.23\times10^{-19}$ & $2.04\times10^{-36}$ \\
$Y_{\text{H}}$                 & $4.78\times10^{-15}$ & $2.32\times10^{-28}$ \\
$Y_{\text{O}}$                 & $1.52\times10^{-13}$ & $4.09\times10^{-26}$ \\
$Y_{\text{H}\text{O}_2}$       & $4.14\times10^{-16}$ & $4.11\times10^{-31}$ \\
$Y_{\text{H}\text{C}\text{O}}$ & $5.70\times10^{-17}$ & $1.05\times10^{-32}$ \\
$Y_{\text{C}\text{H}_2(\text{S})}$ & $2.79\times10^{-19}$ & $3.61\times10^{-37}$ \\
$Y_{\text{C}_2\text{H}_2}$     & $7.39\times10^{-15}$ & $5.05\times10^{-28}$ \\
$Y_{\text{C}_2\text{H}_4}$     & $8.49\times10^{-15}$ & $1.89\times10^{-28}$ \\
$Y_{\text{C}_2\text{H}_6}$     & $1.04\times10^{-14}$ & $3.88\times10^{-28}$ \\
$Y_{\text{C}_2\text{H}_3}$     & $8.14\times10^{-18}$ & $1.40\times10^{-34}$ \\
$Y_{\text{C}\text{H}_2\text{C}\text{O}}$ & $1.25\times10^{-15}$ & $3.71\times10^{-30}$ \\
$\sum$ other species&$1.09\times10^{-9}$ &$1.67\times10^{-18}$\\
$V$&$1.83\times10^{-6}$ &$8.27\times10^{-12}$\\
\hline
\end{tabular}
\caption{Average weighted mean square error ($E_W$) and its variance for each species mass fraction and the velocity predicted by the state-completing FNO across $2,625$ methane-air test flow cases. The averages and variances were computed using Equations~\ref{wmse} and~\ref{var}. Values for $V$ are in $\mathrm{m}^2\,\mathrm{s}^{-2}$; all others are dimensionless.}
\label{table:wmse_methane}
\end{center}
\end{table}
\begin{table}[h!]
\begin{center}\scriptsize
\vskip -0.05in
\begin{tabular}{c c c}
\textbf{Variable} & \textbf{$E_{\max}$} & \textbf{$\mathrm{Var}(E_{\max})$} \\
\hline
$Y_{\text{H}_2\text{O}}$       & $9.41\times10^{-4}$ & $2.83\times10^{-7}$ \\
$Y_{\text{C}\text{O}_2}$       & $1.25\times10^{-3}$ & $3.43\times10^{-7}$ \\
$Y_{\text{C}\text{O}}$         & $7.07\times10^{-4}$ & $1.49\times10^{-7}$ \\
$Y_{\text{H}_2}$               & $2.97\times10^{-5}$ & $2.60\times10^{-10}$ \\
$Y_{\text{O}\text{H}}$         & $8.30\times10^{-5}$ & $2.82\times10^{-9}$ \\
$Y_{\text{C}\text{H}_3}$       & $3.07\times10^{-5}$ & $4.43\times10^{-10}$ \\
$Y_{\text{C}\text{H}_2\text{O}}$ & $1.66\times10^{-5}$ & $7.07\times10^{-11}$ \\
$Y_{\text{H}_2\text{O}_2}$     & $1.12\times10^{-6}$ & $2.54\times10^{-13}$ \\
$Y_{\text{C}\text{H}_3\text{O}\text{H}}$ & $2.32\times10^{-6}$ & $1.57\times10^{-12}$ \\
$Y_{\text{C}\text{H}_2\text{O}\text{H}}$ & $1.62\times10^{-7}$ & $1.45\times10^{-14}$ \\
$Y_{\text{H}}$                 & $5.61\times10^{-6}$ & $1.69\times10^{-11}$ \\
$Y_{\text{O}}$                 & $4.48\times10^{-5}$ & $6.52\times10^{-10}$ \\
$Y_{\text{H}\text{O}_2}$       & $3.49\times10^{-6}$ & $2.97\times10^{-12}$ \\
$Y_{\text{H}\text{C}\text{O}}$ & $1.31\times10^{-6}$ & $6.80\times10^{-13}$ \\
$Y_{\text{C}\text{H}_2(\text{S})}$ & $8.34\times10^{-8}$ & $5.30\times10^{-15}$ \\
$Y_{\text{C}_2\text{H}_2}$     & $1.25\times10^{-5}$ & $3.44\times10^{-10}$ \\
$Y_{\text{C}_2\text{H}_4}$     & $1.39\times10^{-5}$ & $1.78\times10^{-10}$ \\
$Y_{\text{C}_2\text{H}_6}$     & $1.91\times10^{-5}$ & $1.76\times10^{-10}$ \\
$Y_{\text{C}_2\text{H}_3}$     & $6.09\times10^{-7}$ & $4.21\times10^{-13}$ \\
$Y_{\text{C}\text{H}_2\text{C}\text{O}}$ & $4.76\times10^{-6}$ & $2.43\times10^{-11}$ \\
$\sum$ other species&$1.96\times10^{-3}$ &$8.86\times10^{-7}$\\
$V$&$7.59\times10^{-2}$ &$2.60\times10^{-3}$\\
\hline
\end{tabular}
\caption{Average maximum pointwise error ($E_{\max}$) and its variance for each species mass fraction and the velocity predicted by the state-completing FNO across $2,625$ methane-air test flow cases. The averages and variances were computed using Equations~\ref{maxError} and~\ref{varMax}. Values for $V$ are in $\mathrm{m}\,\mathrm{s}^{-1}$; all others are dimensionless.}
\label{table:maxerror_methane}
\end{center}
\end{table}

\begin{figure}[h!]
\centering
\includegraphics[width=0.6\textwidth]{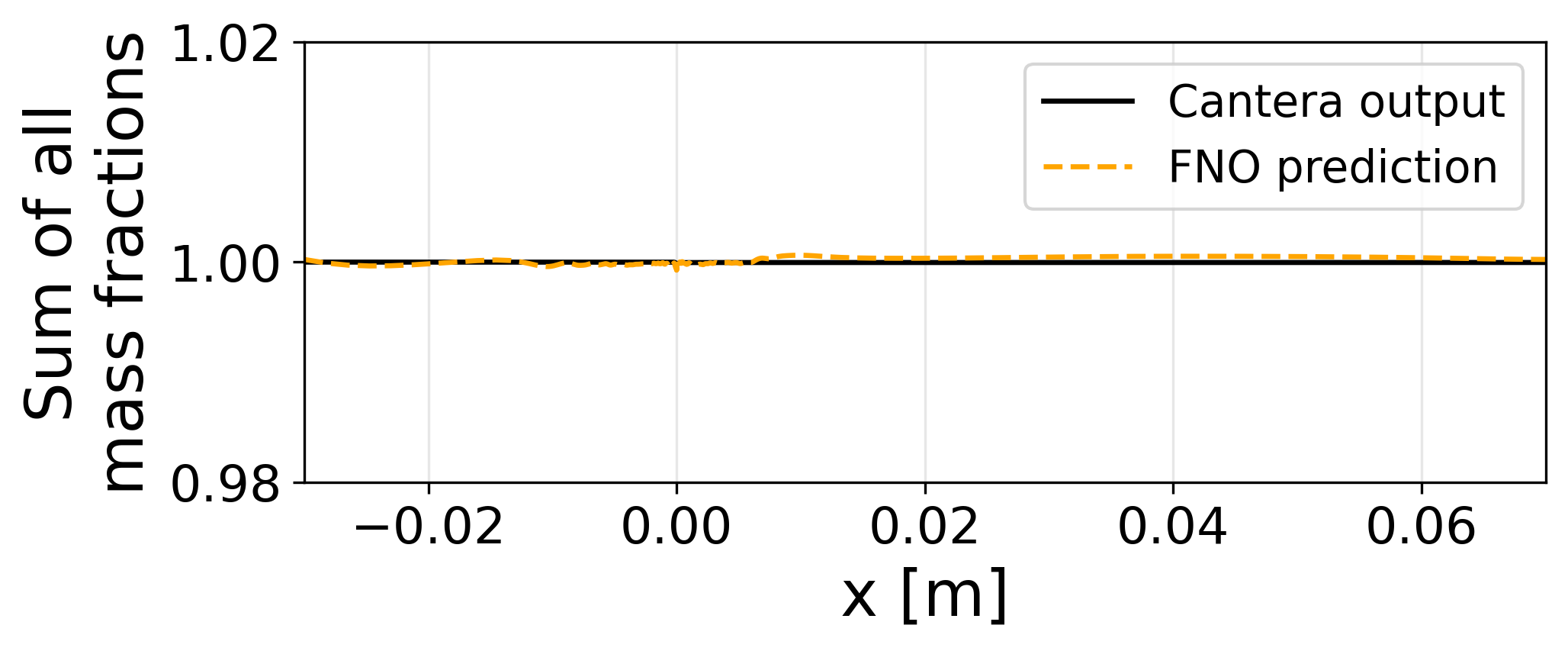}
\caption{Average sum of mass fractions over $2,625$ test flow cases for the methane flame configuration. The $x$-axis origin is the recentering anchor $x_{50\%}$, the location where the temperature reaches $50\%$ of its total rise.}
\label{FNO_sumY_methane}
\end{figure}

As in the hydrogen case, an additional way of testing the performance of the model for all predicted species is by looking at the sum of the computed mass fractions using Equation \ref{sum_massFrac}. Figure \ref{FNO_sumY_methane} shows the average sum of the species for all $2,625$ test cases. This figure shows that, as in the hydrogen case, the accurate predictions of the FNO for all species help the model satisfy the physical condition that Equation \ref{sum_massFrac} equals one at every point in space, even though this requirement is not explicitly enforced in the model.

Despite the substantially larger state size for the methane case ($53$ species and $325$ reactions compared to $9$ species and $25$ reactions for hydrogen), the FNO architecture, using the same number of convolutional neurons and Fourier modes as in the hydrogen case, is still able to produce results with consistently small prediction errors. This result means that this FNO framework is likely to scale well to other, even more complex hydrocarbon mechanisms.

\subsection{Spatially-completing FNO results for a hydrogen flame\label{subsec:spatially_comp_FNO_results}} 
\begin{figure}[h!]
\centering
\includegraphics[width=0.6\textwidth]{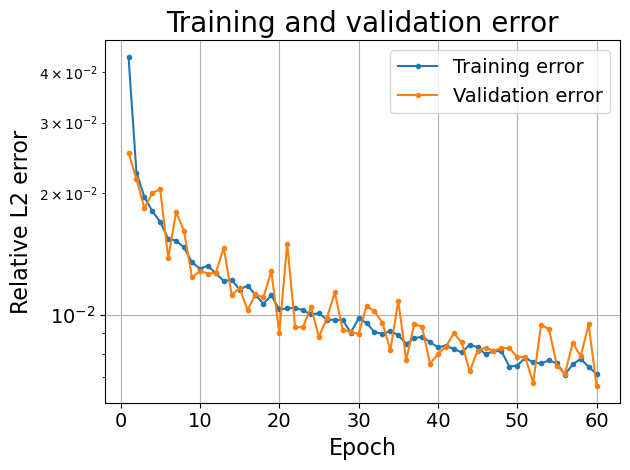}
\caption{Training errors for spatial completion of a hydrogen-air flame measured using Equation \ref{eq:loss}.}
\label{spatial_train_err_hyd}
\end{figure}

As described in Section \ref{subsec:spatially_comp_FNO}, this section looks at the spatial completion of a hydrogen-air flame given information about the species before $x_{10\%}$. Figure \ref{spatial_train_err_hyd} shows the convergence of the training and validation errors during the training of the model. This figure demonstrates that, overall, both the training and validation curve decrease as the number of epochs increases, with both of them converging into errors with similar magnitudes after $60$ epochs. 

\begin{figure}[h!]
\centering
\includegraphics[width=0.6\textwidth]{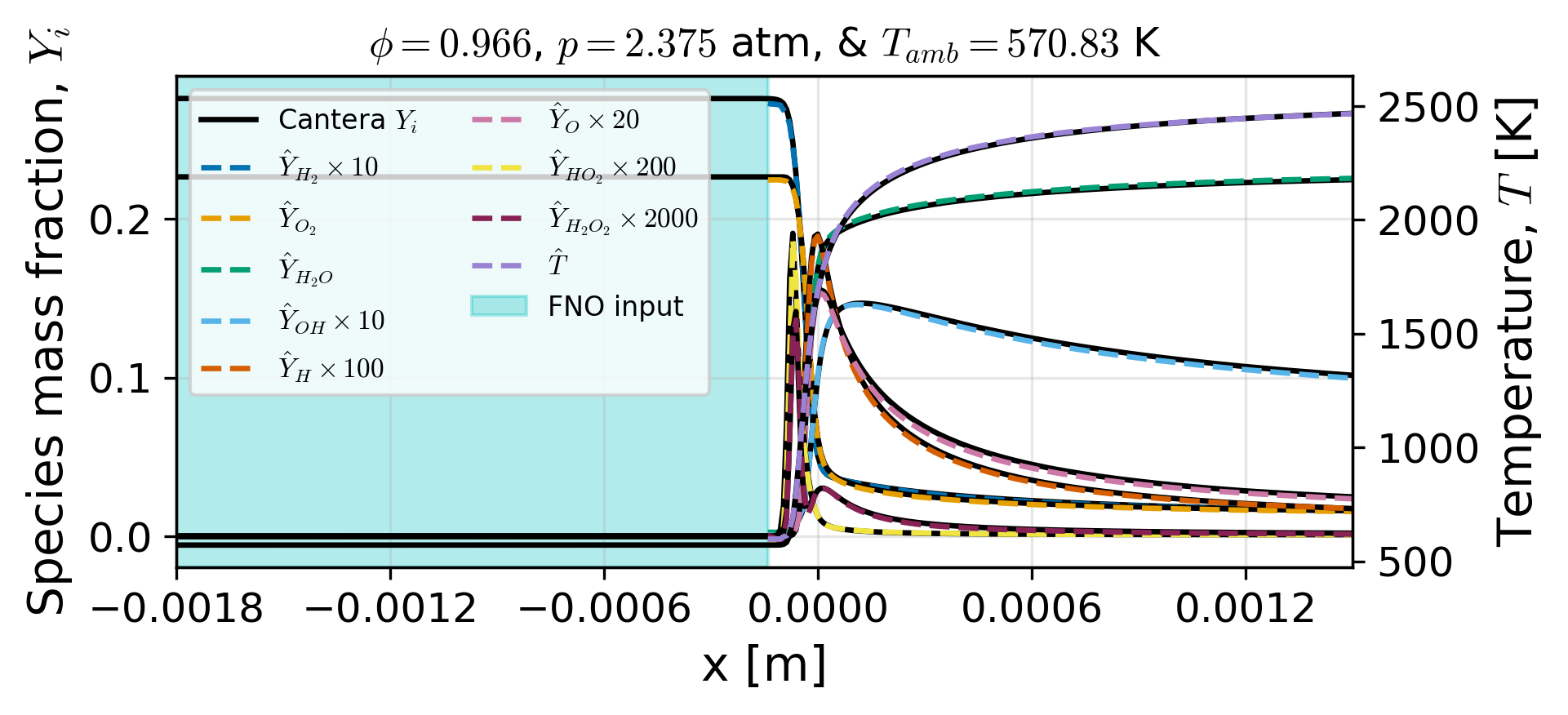}
\caption{Spatially-completing FNO temperature and mass fraction predictions for a single representative test case of a hydrogen flame. The shaded region marks the upstream state supplied to the FNO. The $x$-axis origin is the recentering anchor $x_{50\%}$, the location where the temperature reaches $50\%$ of its total rise.}
\label{FNO_spatial_pred}
\end{figure}

Figure \ref{FNO_spatial_pred} shows the spatial predictions obtained by implementing the architecture described in Section \ref{subsec:spatially_comp_FNO} on hydrogen flame data. In this figure, the region shaded in blue corresponds to the spatial region used as the input to the FNO (the region ahead of the average $x_{10\%}$), and the dotted lines represent the species mass fraction predictions after this point. The $x$-axis is centered around the $x_{50\%}$ location for this particular flow case.

The error metrics defined in Equations \ref{wmse}, \ref{maxError}, \ref{var}, and \ref{varMax} were computed for $2,625$ test cases for the spatial region after $x_{10\%}$. Table~\ref{table:wmse_spatial} and Table~\ref{table:maxerror_spatial} show the result of these metrics for $E_W$ and $E_{\max}$, respectively. Once again, these metrics show that there is good global and local agreement between the predicted profiles and the outputs from Cantera. This indicates that the FNO successfully learns a functional relationship between upstream gradients and the subsequent evolution of species mass fractions and temperature to be able to accurately reconstruct the downstream thermochemical composition of the flame.

\begin{table}[h!]
\begin{center}\scriptsize
\vskip -0.05in
\begin{tabular}{c c c}
\textbf{Variable} & \textbf{$E_W$} & \textbf{$\mathrm{Var}(E_W)$} \\
\hline
$T$                            & $3.39\times10^{-2}$ & $4.28\times10^{-3}$ \\
$Y_{\text{H}_2}$               & $3.85\times10^{-12}$ & $4.93\times10^{-23}$ \\
$Y_{\text{O}_2}$               & $2.86\times10^{-10}$ & $1.05\times10^{-19}$ \\
$Y_{\text{H}_2\text{O}}$       & $3.47\times10^{-10}$ & $1.52\times10^{-19}$ \\
$Y_{\text{O}\text{H}}$         & $1.38\times10^{-11}$ & $1.40\times10^{-21}$ \\
$Y_{\text{N}_2}$               & $9.73\times10^{-12}$ & $1.43\times10^{-22}$ \\
$Y_{\text{H}}$                 & $1.48\times10^{-13}$ & $7.55\times10^{-26}$ \\
$Y_{\text{O}}$                 & $8.74\times10^{-13}$ & $1.75\times10^{-24}$ \\
$Y_{\text{H}\text{O}_2}$       & $1.42\times10^{-14}$ & $1.57\times10^{-27}$ \\
$Y_{\text{H}_2\text{O}_2}$     & $4.92\times10^{-16}$ & $2.28\times10^{-30}$ \\
$V$                            & $4.19\times10^{-5}$ & $4.24\times10^{-9}$ \\
\hline
\end{tabular}
\caption{Average weighted mean square error ($E_W$) and its variance for the temperature, each species mass fraction, and the velocity predicted by the spatially-completing FNO across $2,625$ hydrogen-air test flow cases. The averages and variances were computed using Equations~\ref{wmse} and~\ref{var}. Values for $T$ are in $\mathrm{K}^2$ and values for $V$ are in $\mathrm{m}^2\,\mathrm{s}^{-2}$; all others are dimensionless.}
\label{table:wmse_spatial}
\end{center}
\end{table}

\begin{table}[h!]
\begin{center}\scriptsize
\vskip -0.05in
\begin{tabular}{c c c}
\textbf{Variable} & \textbf{$E_{\max}$} & \textbf{$\mathrm{Var}(E_{\max})$} \\
\hline
$T$                            & $4.23\times10^{1}$ & $1.91\times10^{3}$ \\
$Y_{\text{H}_2}$               & $6.24\times10^{-4}$ & $4.44\times10^{-7}$ \\
$Y_{\text{O}_2}$               & $3.58\times10^{-3}$ & $1.10\times10^{-5}$ \\
$Y_{\text{H}_2\text{O}}$       & $4.07\times10^{-3}$ & $1.55\times10^{-5}$ \\
$Y_{\text{O}\text{H}}$         & $2.59\times10^{-4}$ & $4.30\times10^{-8}$ \\
$Y_{\text{N}_2}$               & $7.63\times10^{-4}$ & $1.07\times10^{-7}$ \\
$Y_{\text{H}}$                 & $5.83\times10^{-5}$ & $2.97\times10^{-9}$ \\
$Y_{\text{O}}$                 & $1.78\times10^{-4}$ & $2.76\times10^{-8}$ \\
$Y_{\text{H}\text{O}_2}$       & $8.86\times10^{-5}$ & $2.23\times10^{-8}$ \\
$Y_{\text{H}_2\text{O}_2}$     & $1.34\times10^{-5}$ & $6.50\times10^{-10}$ \\
$V$                            & $4.08\times10^{-1}$ & $1.30\times10^{-1}$ \\
\hline
\end{tabular}
\caption{Average maximum pointwise error ($E_{\max}$) and its variance for the temperature, each species mass fraction, and the velocity predicted by the spatially-completing FNO across $2,625$ hydrogen-air test flow cases. The averages and variances were computed using Equations~\ref{maxError} and~\ref{varMax}. Values for $T$ are in $\mathrm{K}$ and values for $V$ are in $\mathrm{m}\,\mathrm{s}^{-1}$; all others are dimensionless.}
\label{table:maxerror_spatial}
\end{center}
\end{table}

\subsection{Spatially-completing FNO results for a methane flame\label{subsec:spatially_comp_FNO_methane}} 

\begin{figure}[h!]
\centering
\includegraphics[width=0.6\textwidth]{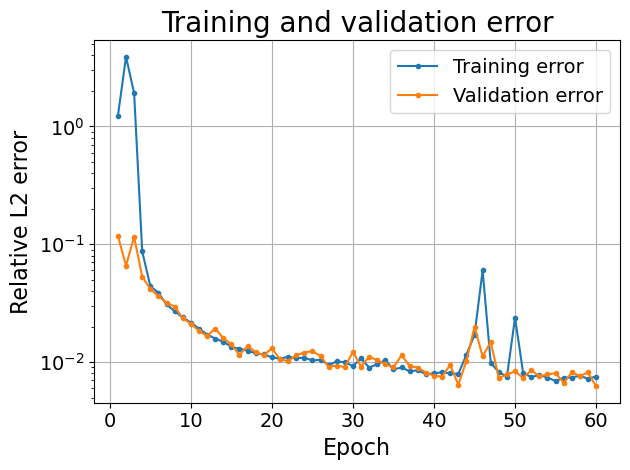}
\caption{Training errors for spatial completion of a methane-air flame measured using Equation \ref{eq:loss}.}
\label{spatial_train_err_methane}
\end{figure}

This section looks at performing a similar spatial-completion task as Section \ref{subsec:spatially_comp_FNO_results}, but for the more complex case of methane combustion. The training and validation errors during the training process are shown in Figure \ref{spatial_train_err_methane}.

\begin{figure}[h!]
\centering
\includegraphics[width=0.6\textwidth]{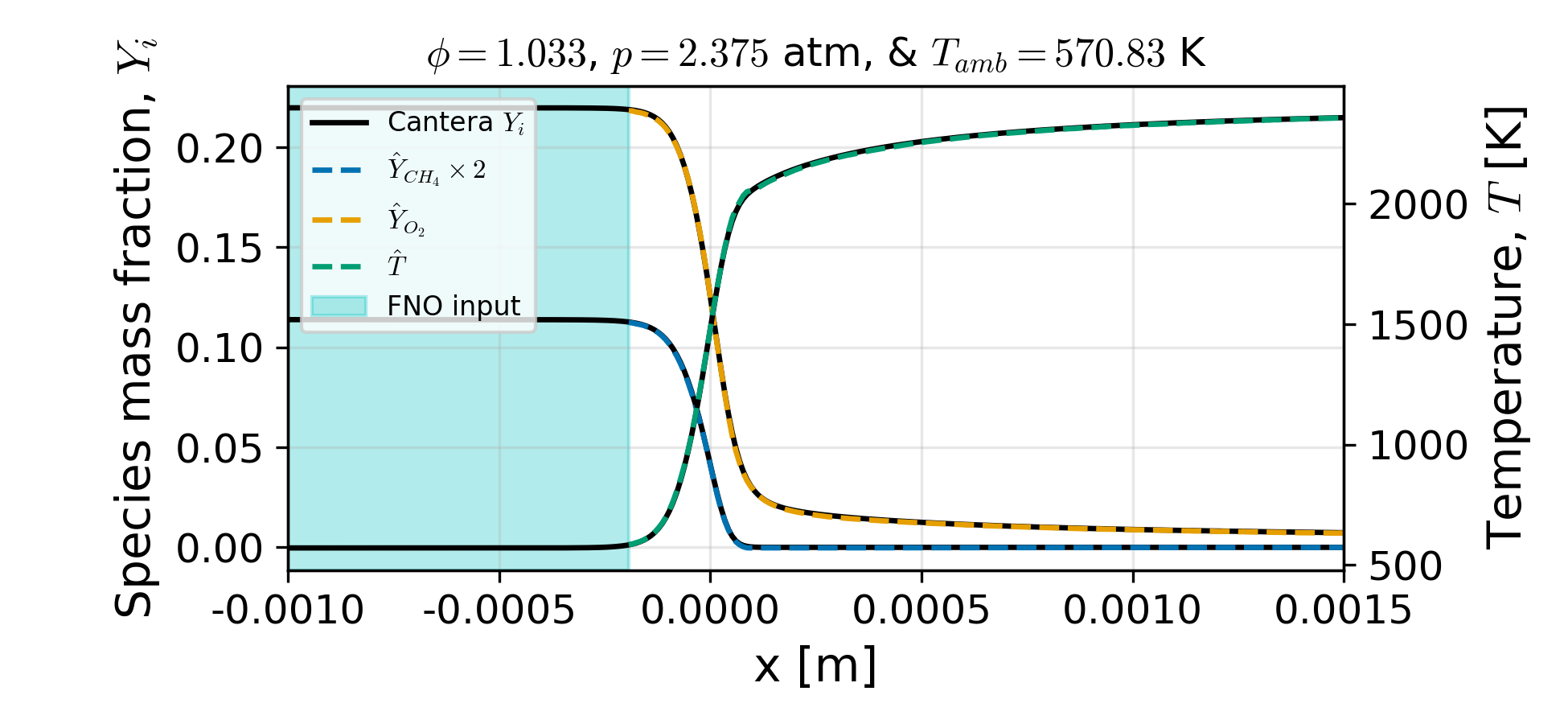}
\caption{Spatially-completing FNO predictions of the reactant mass fractions and temperature for a single representative test case of a methane flame. The shaded region marks the upstream state supplied to the FNO. The $x$-axis origin is the recentering anchor $x_{50\%}$, the location where the temperature reaches $50\%$ of its total rise.}
\label{FNO_spatial_pred_ch4_reactants}
\end{figure}

\begin{figure}[h!]
\centering
\includegraphics[width=0.6\textwidth]{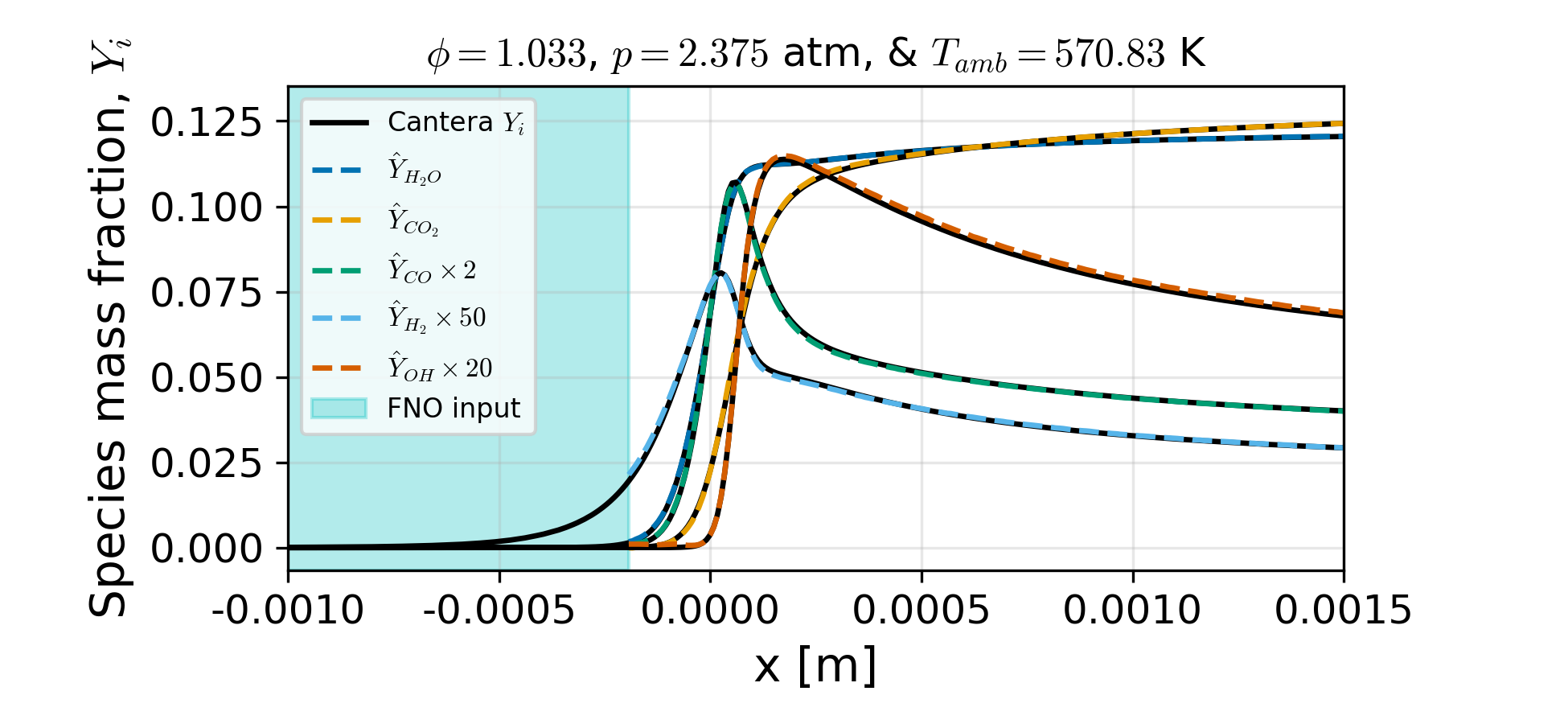}
\caption{Spatially-completing FNO mass fraction predictions for a set of major combustion species for a single representative test case of a methane flame. The shaded region marks the upstream state supplied to the FNO. The $x$-axis origin is the recentering anchor $x_{50\%}$, the location where the temperature reaches $50\%$ of its total rise.}
\label{FNO_spatial_pred_ch4_major}
\end{figure}

\begin{figure}[h!]
\centering
\includegraphics[width=0.6\textwidth]{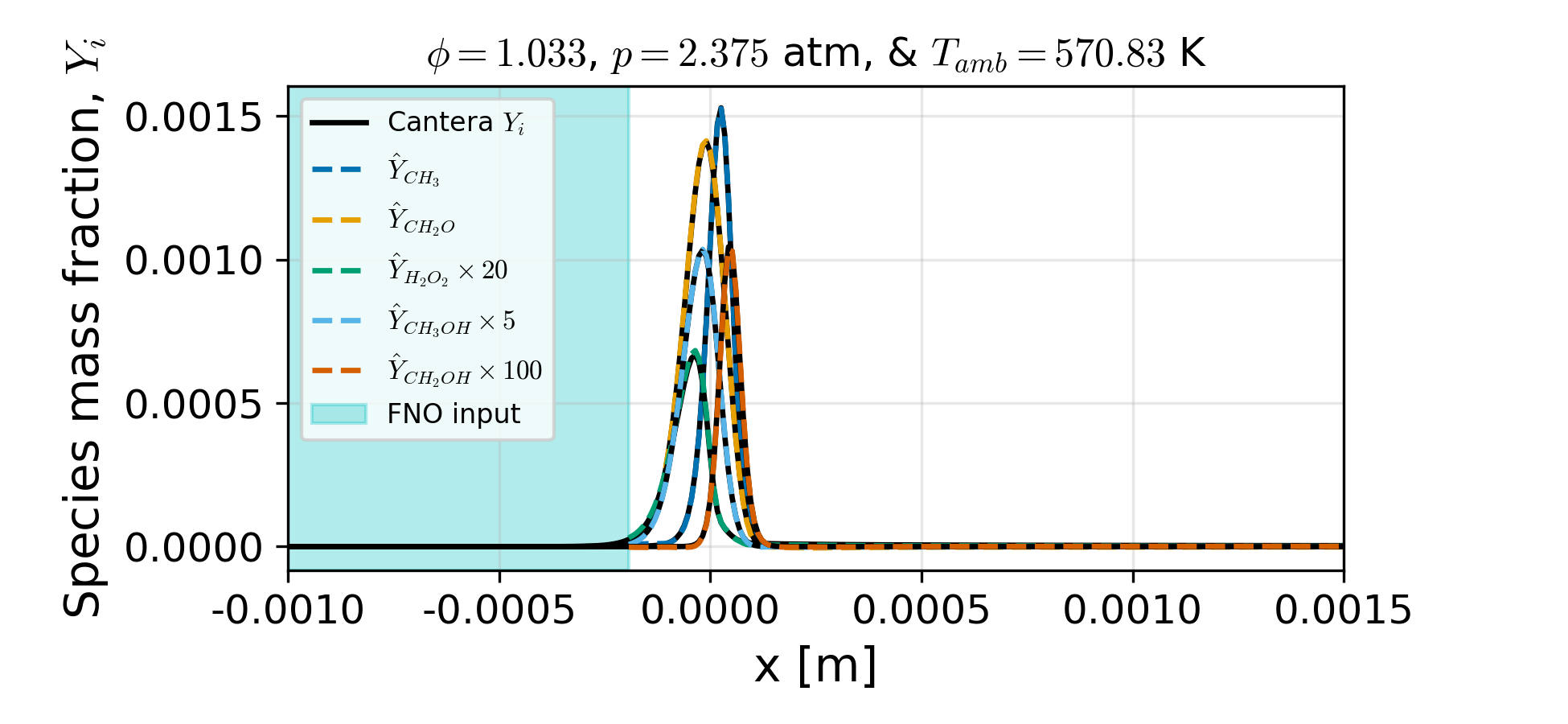}
\caption{Spatially-completing FNO mass fraction predictions for a set of intermediate combustion species for a single representative test case of a methane flame. The shaded region marks the upstream state supplied to the FNO. The $x$-axis origin is the recentering anchor $x_{50\%}$, the location where the temperature reaches $50\%$ of its total rise.}
\label{FNO_spatial_pred_ch4_int1}
\end{figure}

\begin{figure}[h!]
\centering
\includegraphics[width=0.6\textwidth]{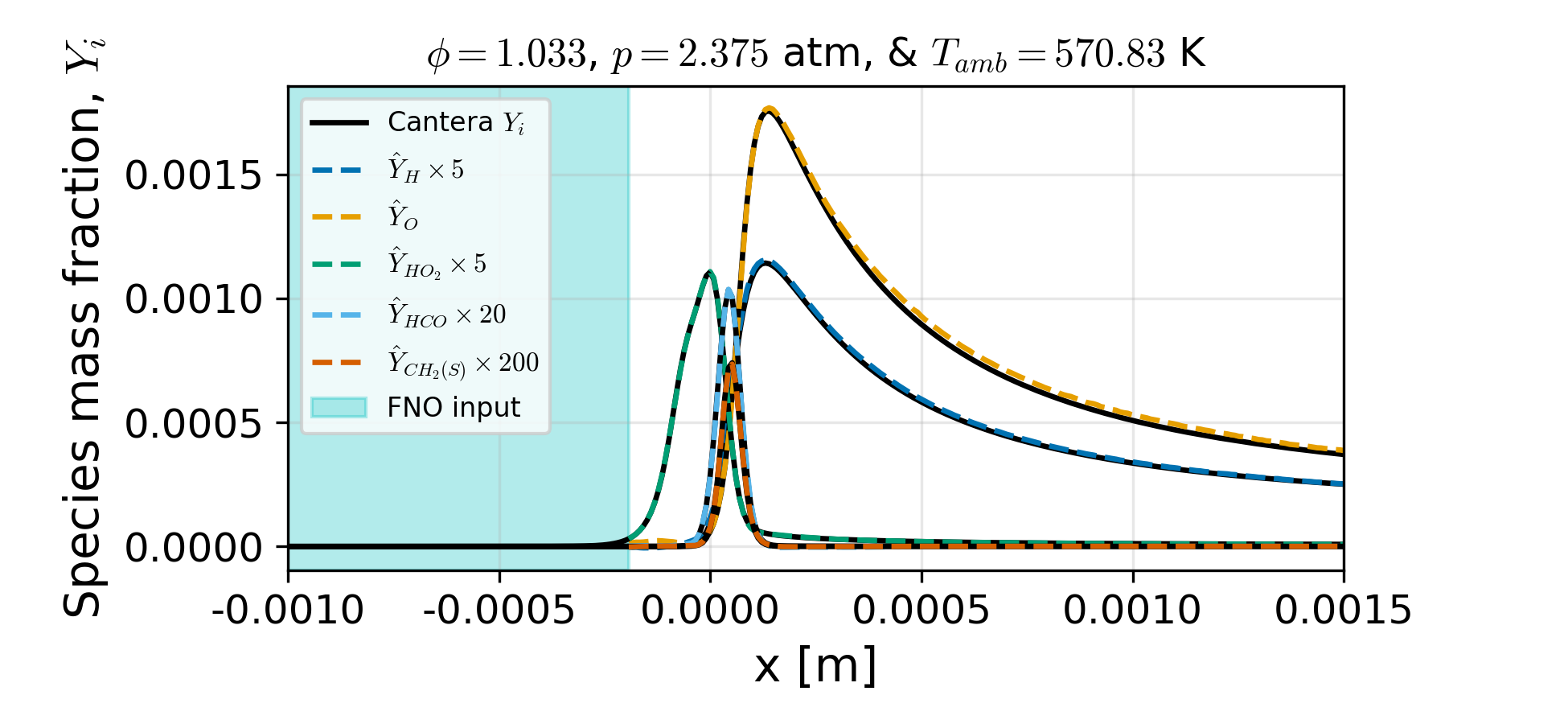}
\caption{Spatially-completing FNO mass fraction predictions for a set of additional intermediate combustion species for a single representative test case of a methane flame. The shaded region marks the upstream state supplied to the FNO. The $x$-axis origin is the recentering anchor $x_{50\%}$, the location where the temperature reaches $50\%$ of its total rise.}
\label{FNO_spatial_pred_ch4_int2}
\end{figure}

\begin{figure}[h!]
\centering
\includegraphics[width=0.6\textwidth]{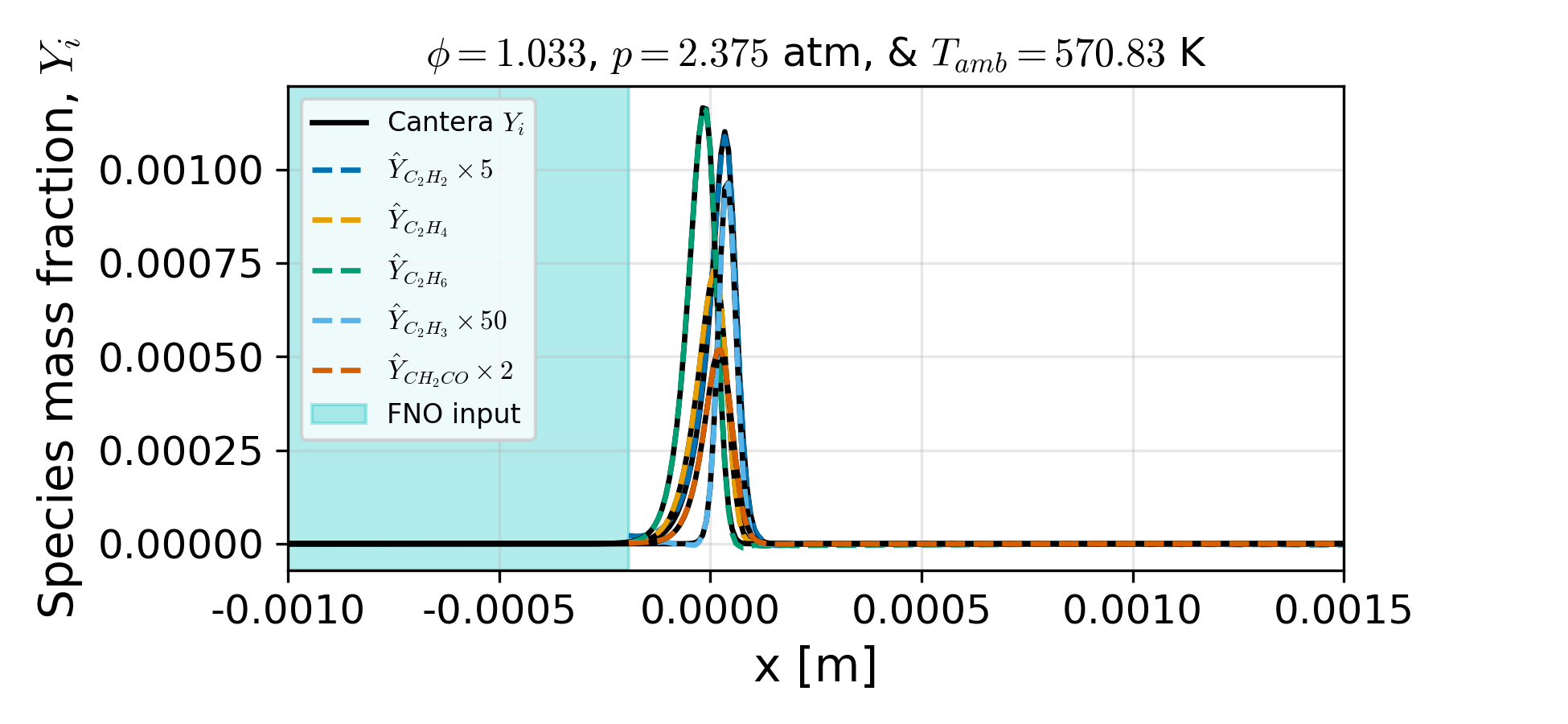}
\caption{Spatially-completing FNO mass fraction predictions for a set of trace combustion species for a single representative test case of a methane flame. The shaded region marks the upstream state supplied to the FNO. The $x$-axis origin is the recentering anchor $x_{50\%}$, the location where the temperature reaches $50\%$ of its total rise.}
\label{FNO_spatial_pred_ch4_trace}
\end{figure}

Figures~\ref{FNO_spatial_pred_ch4_reactants}--\ref{FNO_spatial_pred_ch4_trace} show the FNO predictions for a single representative case for a select set of species. As in Section \ref{subsec:state_comp_FNO_res_methane}, the plots of the predictions are grouped by their role in the mechanism, showing the reactants together with temperature, the major products, two sets of intermediates, and the trace hydrocarbon species.

Across all five groups the predicted profiles follow the Cantera solutions closely over the whole of the predicted region. As with the hydrogen case, the error metrics defined in Equations \ref{wmse}, \ref{maxError}, \ref{var}, and \ref{varMax} were computed for the spatial region after $x_{10\%}$. Table~\ref{table:wmse_methane_spatial} and Table~\ref{table:maxerror_methane_spatial} show the result of these metrics. 

\begin{table}[h!]
\begin{center}\scriptsize
\vskip -0.05in
\begin{tabular}{c c c}
\textbf{Variable} & \textbf{$E_W$} & \textbf{$\mathrm{Var}(E_W)$} \\
\hline
$T$                            & $7.16\times10^{-2}$ & $3.79\times10^{-2}$ \\
$Y_{\text{C}\text{H}_4}$       & $1.09\times10^{-11}$ & $2.79\times10^{-22}$ \\
$Y_{\text{O}_2}$               & $4.64\times10^{-10}$ & $7.29\times10^{-19}$ \\
$Y_{\text{H}_2\text{O}}$       & $1.26\times10^{-10}$ & $6.73\times10^{-20}$ \\
$Y_{\text{C}\text{O}_2}$       & $2.88\times10^{-10}$ & $9.32\times10^{-20}$ \\
$Y_{\text{C}\text{O}}$         & $1.03\times10^{-10}$ & $5.79\times10^{-20}$ \\
$Y_{\text{H}_2}$               & $7.42\times10^{-13}$ & $1.27\times10^{-24}$ \\
$Y_{\text{O}\text{H}}$         & $2.83\times10^{-12}$ & $2.57\times10^{-23}$ \\
$Y_{\text{C}\text{H}_3}$       & $1.60\times10^{-14}$ & $5.18\times10^{-28}$ \\
$Y_{\text{C}\text{H}_2\text{O}}$ & $5.63\times10^{-15}$ & $2.34\times10^{-28}$ \\
$Y_{\text{H}_2\text{O}_2}$     & $1.66\times10^{-17}$ & $1.41\times10^{-33}$ \\
$Y_{\text{C}\text{H}_3\text{O}\text{H}}$ & $7.73\times10^{-17}$ & $2.43\times10^{-32}$ \\
$Y_{\text{C}\text{H}_2\text{O}\text{H}}$ & $4.48\times10^{-19}$ & $1.17\times10^{-36}$ \\
$Y_{\text{H}}$                 & $2.49\times10^{-15}$ & $3.24\times10^{-29}$ \\
$Y_{\text{O}}$                 & $2.08\times10^{-13}$ & $2.01\times10^{-25}$ \\
$Y_{\text{H}\text{O}_2}$       & $3.40\times10^{-16}$ & $1.57\times10^{-31}$ \\
$Y_{\text{H}\text{C}\text{O}}$ & $2.46\times10^{-17}$ & $2.32\times10^{-33}$ \\
$Y_{\text{C}\text{H}_2(\text{S})}$ & $9.09\times10^{-20}$ & $3.56\times10^{-38}$ \\
$Y_{\text{C}_2\text{H}_2}$     & $1.93\times10^{-15}$ & $2.65\times10^{-29}$ \\
$Y_{\text{C}_2\text{H}_4}$     & $3.68\times10^{-15}$ & $7.66\times10^{-29}$ \\
$Y_{\text{C}_2\text{H}_6}$     & $6.87\times10^{-15}$ & $6.22\times10^{-29}$ \\
$Y_{\text{C}_2\text{H}_3}$     & $3.08\times10^{-18}$ & $4.19\times10^{-35}$ \\
$Y_{\text{C}\text{H}_2\text{C}\text{O}}$ & $4.32\times10^{-16}$ & $1.27\times10^{-30}$ \\
$\sum$ other species&$4.12\times10^{-11}$ &$5.20\times10^{-21}$\\
$V$&$1.33\times10^{-6}$ &$6.89\times10^{-12}$\\
\hline
\end{tabular}
\caption{Average weighted mean square error ($E_W$) and its variance for the temperature, each species mass fraction, and the velocity predicted by the spatially-completing FNO across $2,625$ methane-air test flow cases. The averages and variances were computed using Equations~\ref{wmse} and~\ref{var}. Values for $T$ are in $\mathrm{K}^2$ and values for $V$ are in $\mathrm{m}^2\,\mathrm{s}^{-2}$; all others are dimensionless.}
\label{table:wmse_methane_spatial}
\end{center}
\end{table}
\begin{table}[h!]
\begin{center}\scriptsize
\vskip -0.05in
\begin{tabular}{c c c}
\textbf{Variable} & \textbf{$E_{\max}$} & \textbf{$\mathrm{Var}(E_{\max})$} \\
\hline
$T$                            & $1.44\times10^{1}$ & $4.94\times10^{1}$ \\
$Y_{\text{C}\text{H}_4}$       & $4.34\times10^{-4}$ & $7.42\times10^{-8}$ \\
$Y_{\text{O}_2}$               & $1.49\times10^{-3}$ & $5.15\times10^{-7}$ \\
$Y_{\text{H}_2\text{O}}$       & $7.61\times10^{-4}$ & $1.53\times10^{-7}$ \\
$Y_{\text{C}\text{O}_2}$       & $1.21\times10^{-3}$ & $2.88\times10^{-7}$ \\
$Y_{\text{C}\text{O}}$         & $7.11\times10^{-4}$ & $2.50\times10^{-7}$ \\
$Y_{\text{H}_2}$               & $4.50\times10^{-5}$ & $5.82\times10^{-10}$ \\
$Y_{\text{O}\text{H}}$         & $6.19\times10^{-5}$ & $8.71\times10^{-10}$ \\
$Y_{\text{C}\text{H}_3}$       & $1.94\times10^{-5}$ & $1.69\times10^{-10}$ \\
$Y_{\text{C}\text{H}_2\text{O}}$ & $1.39\times10^{-5}$ & $5.36\times10^{-11}$ \\
$Y_{\text{H}_2\text{O}_2}$     & $7.09\times10^{-7}$ & $1.15\times10^{-13}$ \\
$Y_{\text{C}\text{H}_3\text{O}\text{H}}$ & $1.97\times10^{-6}$ & $7.75\times10^{-13}$ \\
$Y_{\text{C}\text{H}_2\text{O}\text{H}}$ & $1.38\times10^{-7}$ & $8.03\times10^{-15}$ \\
$Y_{\text{H}}$                 & $4.52\times10^{-6}$ & $5.59\times10^{-12}$ \\
$Y_{\text{O}}$                 & $2.89\times10^{-5}$ & $1.71\times10^{-10}$ \\
$Y_{\text{H}\text{O}_2}$       & $2.93\times10^{-6}$ & $2.31\times10^{-12}$ \\
$Y_{\text{H}\text{C}\text{O}}$ & $9.26\times10^{-7}$ & $3.67\times10^{-13}$ \\
$Y_{\text{C}\text{H}_2(\text{S})}$ & $6.66\times10^{-8}$ & $2.29\times10^{-15}$ \\
$Y_{\text{C}_2\text{H}_2}$     & $8.86\times10^{-6}$ & $8.35\times10^{-11}$ \\
$Y_{\text{C}_2\text{H}_4}$     & $1.04\times10^{-5}$ & $5.99\times10^{-11}$ \\
$Y_{\text{C}_2\text{H}_6}$     & $1.35\times10^{-5}$ & $5.70\times10^{-11}$ \\
$Y_{\text{C}_2\text{H}_3}$     & $3.52\times10^{-7}$ & $7.37\times10^{-14}$ \\
$Y_{\text{C}\text{H}_2\text{C}\text{O}}$ & $3.44\times10^{-6}$ & $7.72\times10^{-12}$ \\
$\sum$ other species&$4.86\times10^{-4}$ &$4.84\times10^{-8}$\\
$V$&$6.00\times10^{-2}$ &$1.12\times10^{-3}$\\
\hline
\end{tabular}
\caption{Average maximum pointwise error ($E_{\max}$) and its variance for the temperature, each species mass fraction, and the velocity predicted by the spatially-completing FNO across $2,625$ methane-air test flow cases. The averages and variances were computed using Equations~\ref{maxError} and~\ref{varMax}. Values for $T$ are in $\mathrm{K}$ and values for $V$ are in $\mathrm{m}\,\mathrm{s}^{-1}$; all others are dimensionless.}
\label{table:maxerror_methane_spatial}
\end{center}
\end{table}

Together, these results demonstrate two forms of reconstruction within the sampled premixed-flame conditions. State completion recovers radical, intermediate, and product profiles without supplying those species, while spatial completion recovers downstream structure from upstream profiles extending into the initial temperature rise. The recovered quantities therefore include localized reaction-zone features as well as post-flame composition.

\section{Conclusions\label{sec:conclusions}}

This work demonstrated reconstruction of coupled thermochemical structure from incomplete descriptions of steady premixed flames within the sampled conditions. For hydrogen-air and methane-air flames, FNOs recovered missing state variables and downstream profiles, including localized reaction-zone species. These results support using FNOs to infer unobserved quantities from partial flame profiles.

The state-completing FNO successfully reconstructed the full thermochemical state using only temperature and two major reactant species as inputs, achieving low weighted mean squared errors and maximum pointwise errors across a large test set for both hydrogen-air and methane-air combustion. The spatially-completing FNO demonstrated that it is possible to predict the downstream flame structure using the upstream thermochemical state information as an input, highlighting the potential of operator-learning approaches for reconstructing missing spatial information.

The current implementation is purely data-driven and therefore only implicitly captures conservation and reaction constraints. Future work will introduce explicit physical terms into the loss function and explore coupling strategies that integrate the ROM into CFD and experimental workflows, where only a subset of the thermochemical state is available during transport. 
Sparse diagnostics, inference of inaccessible quantities, and reduced-state modeling are potential applications of these reconstruction capabilities.

\section*{CRediT authorship contribution statement}

{\bf Federico~Rios~Tascon}: Designed research, developed code, analyzed data, developed ML techniques, and wrote the paper. {\bf Ryan~F.~Johnson}: Wrote paper, designed and guided research. {\bf Diego~D.~Ortiz}: Developed techniques and analyzed data. {\bf Peter~J.~Schmid}: Co-principal investigator, designed and guided research. {\bf Beverley~J.~McKeon}: Principal investigator, project management, designed and guided research, project funds.

\section*{Declaration of competing interest}

The authors declare that they have no known competing financial interests or personal relationships that could have appeared to influence the work reported in this paper.

\section*{Acknowledgments}

The support of the Office of Naval Research under grant number N00014-23-1-2646 under Dr. Eric Marineau of the Hypersonic Aerothermodynamics, High-Speed Propulsion and Materials Program of the Office of Naval Research Code 35 is gratefully acknowledged. Ryan Johnson also acknowledges Dr. Marineau for partial support.

\FloatBarrier

\bibliographystyle{elsarticle-num}
\bibliography{references}

@misc{goodwin2016cantera,
  title={Cantera: An Object-oriented Software Toolkit for Chemical Kinetics, Thermodynamics, and Transport Processes},
  author={Goodwin, D.G. and Moffat, H.K. and Speth, R.L.},
  year={2016},
  howpublished={\url{https://cantera.org}}
}

@Misc{ZDV2023,
  author  = {Zhang, Yue and Dong, Wendi and Vandewalle, Laurien and Xu, Rui and Smith, Gregory and Wang, Hai},
  title   = {Foundational {Fuel} {Chemistry} {Model} {Version} 2.0 ({FFCM}-2)},
  journal = {FFCM-2 website},
  url     = "https://web.stanford.edu/group/haiwanglab/FFCM2",
  year    = {2023},
  }

@article{li2021fourierneuraloperatorparametric,
      title={Fourier Neural Operator for Parametric Partial Differential Equations}, 
      author={Zongyi Li and Nikola Kovachki and Kamyar Azizzadenesheli and Burigede Liu and Kaushik Bhattacharya and Andrew Stuart and Anima Anandkumar},
      year={2021},
      journal={arXiv:2010.08895},
      eprint={2010.08895},
      archivePrefix={arXiv},
      primaryClass={cs.LG},
      url={https://arxiv.org/abs/2010.08895}, 
}

@Misc{peters,
  author  = {Peters, Norbert},
  title   = {Combustion Theory - Reference Notes},
  journal = {Combustion Energy Frontier Research Center},
  url     = "https://cefrc.princeton.edu/combustion-summer-school/archived-programs/2010-summer-school/lecture-notes",
  year    = {2010},
  }

@book{poinsot2005theoretical,
  title={Theoretical and Numerical Combustion},
  author={Poinsot, T. and Veynante, D.},
  year={2005},
  publisher={RT Edwards, Inc.}
}

@article{lu2009toward,
  title={Toward accommodating realistic fuel chemistry in large-scale computations},
  author={Lu, T. and Law, C.K.},
  journal={Progress in Energy and Combustion Science},
  volume={35},
  number={2},
  pages={192--215},
  year={2009}
}

@book{Eckbreth1996,
  author    = {Alan C. Eckbreth},
  title     = {Laser Diagnostics for Combustion Temperature and Species},
  edition   = {2},
  publisher = {CRC Press},
  year      = {1996}
}

@book{KohseHoinghaus2002,
  author  = {Kohse-Höinghaus, Katharina and Jeffries, Jay B.},
  title   = {Applied Combustion Diagnostics},
  publisher = {Taylor \& Francis},
  year    = {2002}
}

@article{Alden2011,
  author  = {Ald{\'e}n, Mikael and Omrane, Abdelmajid and Richter, Mattias and Svanberg, Sune},
  title   = {Visualization and understanding of combustion processes using laser-based techniques},
  journal = {Proceedings of the Combustion Institute},
  volume  = {33},
  number  = {1},
  pages   = {69--97},
  year    = {2011}
}

@article{esposito2011skeletal,
  title={Skeletal reaction models based on principal component analysis: Application to ethylene--air ignition, propagation, and extinction phenomena},
  author={Esposito, G and Chelliah, HK},
  journal={Combustion and Flame},
  volume={158},
  number={3},
  pages={477--489},
  year={2011},
  publisher={Elsevier}
}

@article{xu2018physics,
  title={A physics-based approach to modeling real-fuel combustion chemistry--{II.} {Reaction kinetic models of jet and rocket fuels}},
  author={Xu, Rui and Wang, Kun and Banerjee, Sayak and Shao, Jiankun and Parise, Tom and Zhu, Yangye and Wang, Shengkai and Movaghar, Ashkan and Lee, Dong Joon and Zhao, Runhua and others},
  journal={Combustion and Flame},
  volume={193},
  pages={520--537},
  year={2018},
  publisher={Elsevier}
}

@article{lu2005directed,
  title={A directed relation graph method for mechanism reduction},
  author={Lu, Tianfeng and Law, Chung K},
  journal={Proceedings of the Combustion Institute},
  volume={30},
  number={1},
  pages={1333--1341},
  year={2005},
  publisher={Elsevier}
}

@inproceedings{Boj23,
  author = {Bojko, B. T. and Kessler, D. A. and Johnson, R. F.},
  title = {Flamelet Progress Variable Dynamic Source Term and Energy (FPV-DynaSTE) Approach for Compressible Reacting Flows with Heat Loss},
  booktitle = {AIAA Scitech Forum},
  year = {2024}
}

@article{GONZALEZJUEZ201726,
title = {Advances and challenges in modeling high-speed turbulent combustion in propulsion systems},
journal = {Progress in Energy and Combustion Science},
volume = {60},
pages = {26-67},
year = {2017},
issn = {0360-1285},
doi = {10.1016/j.pecs.2016.12.003},
url = {https://www.sciencedirect.com/science/article/pii/S0360128515300563},
author = {Esteban D. Gonzalez-Juez and Alan R. Kerstein and R. Ranjan and S. Menon}
}

@article{ihme2005prediction,
  title={Prediction of local extinction and re-ignition effects in non-premixed turbulent combustion using a flamelet/progress variable approach},
  author={Ihme, Matthias and Cha, Chong M and Pitsch, Heinz},
  journal={Proceedings of the Combustion Institute},
  volume={30},
  number={1},
  pages={793--800},
  year={2005},
  publisher={Elsevier}
}

@book{oran2001numerical,
  title={Numerical Simulation of Reactive Flow},
  author={Oran, E.S. and Boris, J.P.},
  isbn={9780521581752},
  lccn={99059887},
  url={https://books.google.com/books?id=6KzA7QBvoNwC},
  year={2001},
  publisher={Cambridge University Press}
}

@article{KNUDSEN2009678,
title = {A general flamelet transformation useful for distinguishing between premixed and non-premixed modes of combustion},
journal = {Combustion and Flame},
volume = {156},
number = {3},
pages = {678-696},
year = {2009},
issn = {0010-2180},
doi = {10.1016/j.combustflame.2008.10.021},
url = {https://www.sciencedirect.com/science/article/pii/S0010218008003246},
author = {E. Knudsen and H. Pitsch},
}

@article{PIERCE_MOIN_2004, title={Progress-variable approach for large-eddy simulation of non-premixed turbulent combustion}, volume={504}, DOI={10.1017/S0022112004008213}, journal={Journal of Fluid Mechanics}, author={Pierce, Charles D. and Moin, Parviz}, year={2004}, pages={73–97}}

@article{PITSCH19981057,
title = {Unsteady flamelet modeling of turbulent hydrogen-air diffusion flames},
journal = {Symposium (International) on Combustion},
volume = {27},
number = {1},
pages = {1057-1064},
year = {1998},
note = {{Twenty-Seventh Symposium (International) on Combustion Volume One}},
issn = {0082-0784},
doi = {10.1016/S0082-0784(98)80506-7},
url = {https://www.sciencedirect.com/science/article/pii/S0082078498805067},
author = {H. Pitsch and M. Chen and N. Peters},
}

@article{PETERS1984319,
title = {Laminar diffusion flamelet models in non-premixed turbulent combustion},
journal = {Progress in Energy and Combustion Science},
volume = {10},
number = {3},
pages = {319-339},
year = {1984},
issn = {0360-1285},
doi = {10.1016/0360-1285(84)90114-X},
url = {https://www.sciencedirect.com/science/article/pii/036012858490114X},
author = {N. Peters},
}

@article{Clavin_Williams_1979, title={Theory of premixed-flame propagation in large-scale turbulence}, volume={90}, DOI={10.1017/S002211207900241X}, number={3}, journal={Journal of Fluid Mechanics}, author={Clavin, P. and Williams, F. A.}, year={1979}, pages={589–604}}

@misc{gri30,
  author       = {Smith, G. P. and Golden, D. M. and Frenklach, M. and
                  Moriarty, N. W. and Eiteneer, B. and Goldenberg, M. and
                  Bowman, C. T. and Hanson, R. K. and Song, S. and
                  Gardiner, W. C. and Lissianski, V. V. and Qin, Z.},
  title        = {{GRI-Mech} 3.0},
  howpublished = {\url{http://combustion.berkeley.edu/gri-mech/}},
  year         = {1999},
  note         = {Accessed: 2024}
}

@misc{kingma2017adammethodstochasticoptimization,
      title={Adam: A Method for Stochastic Optimization}, 
      author={Diederik P. Kingma and Jimmy Ba},
      year={2017},
      eprint={1412.6980},
      archivePrefix={arXiv},
      primaryClass={cs.LG},
      url={https://arxiv.org/abs/1412.6980}, 
}

@inproceedings{paszke2019pytorch,
  author    = {Adam Paszke and Sam Gross and Francisco Massa and Adam Lerer and
               James Bradbury and Gregory Chanan and Trevor Killeen and Zeming Lin and
               Natalia Gimelshein and Luca Antiga and Alban Desmaison and Andreas K{\"o}pf and
               Edward Yang and Zachary DeVito and Martin Raison and Alykhan Tejani and
               Sasank Chilamkurthy and Benoit Steiner and Lu Fang and Junjie Bai and
               Soumith Chintala},
  title     = {{PyTorch}: An Imperative Style, High-Performance Deep Learning Library},
  booktitle = {Advances in Neural Information Processing Systems (NeurIPS)},
  year      = {2019}
}

@article{weng2025extendedFNO,
  author = {Weng, Yuting and Li, Han and Zhang, Hao and Chen, Zhi X. and Zhou, Dezhi},
  title = {Extended {Fourier Neural Operators} to learn stiff chemical kinetics under unseen conditions},
  journal = {Combustion and Flame},
  volume = {272},
  pages = {113847},
  year = {2025},
  doi = {10.1016/j.combustflame.2024.113847}
}

@article{wu2026resolutionInvariant,
  author = {Wu, Sipei and Liang, Wenkai and Luo, Kai Hong},
  title = {Resolution-invariant temperature and species field reconstruction using factorized {Fourier} neural operator ({F-FNO}) for turbulent jet flames},
  journal = {Combustion and Flame},
  volume = {283},
  pages = {114591},
  year = {2026},
  doi = {10.1016/j.combustflame.2025.114591}
}

@article{vanOijen2000FGM,
  author = {van Oijen, J. A. and de Goey, L. P. H.},
  title = {Modelling of premixed laminar flames using flamelet-generated manifolds},
  journal = {Combustion Science and Technology},
  volume = {161},
  number = {1},
  pages = {113--137},
  year = {2000},
  doi = {10.1080/00102200008935814}
}

\end{document}